\documentclass[11pt]{article}
\usepackage[margin=1in]{geometry}
\usepackage{amsmath,amssymb,amsthm}
\usepackage{microtype}
\usepackage{graphicx}
\usepackage[hidelinks]{hyperref}
\usepackage{graphicx}
\usepackage{color}
\newtheorem{proposition}{Proposition}
\newtheorem{theorem}{Theorem}
\newtheorem{corollary}{Corollary}
\theoremstyle{definition}
\newtheorem{definition}{Definition}
\theoremstyle{definition}
\newtheorem{note}{Note}
\theoremstyle{plain}

\usepackage{lmodern,microtype,amsmath,amssymb,amsthm,mathtools}
\usepackage{graphicx,booktabs,array,longtable,siunitx,listings,xcolor}
\usepackage{tikz}
\usetikzlibrary{arrows.meta}

\newcommand{\surface}{\mathcal Z}
\newcommand{\ggra}{g_{\mathrm{grav}}}
\newcommand{\velphi}{\frac{u_\phi}{R\sin\theta}}
\newcommand{\veltheta}{\frac{u_\theta}{R}}
\newcommand{\dd}{\,\mathrm d}

\newcommand{\code}[1]{\texttt{\detokenize{#1}}}

\newcommand{\dpar}[2]{\dfrac{\partial #1}{\partial #2}}

\definecolor{DarkGreen}{rgb}{0.1,0.6,0.0}

\title{Shallow water equations on n-dimensional Riemannian manifolds: a kinetic approach}
\author{Gerardo Hern\'andez-Due\~nas\\[0.4em]
\small Instituto de Matem\'aticas, Universidad Nacional Aut\'onoma de M\'exico\\
\small Blvd. Juriquilla 3001, Quer\'etaro, Mexico\\
\small \href{mailto:hernandez@im.unam.mx}{\texttt{hernandez@im.unam.mx}}}
\date{}
\begin{document}
\nocite{mueller1992,arpaia2022,montoya2026,perthame2001,mangeney2003,audusse2011,audusse2016,love2012,gonzalez2026}
\maketitle
\begin{abstract}
We develop an intrinsic kinetic moment closure for shallow-water equations on smooth manifolds of arbitrary dimension equipped with prescribed, possibly time-dependent Riemannian metrics. Hamiltonian transport on the cotangent bundle and metric-dependent momentum integration yield balance laws with hydrostatic pressure, bottom topography, a skew-adjoint force, and dissipative bottom friction. An explicit compactly supported Gibbs equilibrium satisfies the required mass, momentum, and tensor second-moment conditions. We derive the quasilinear formulation and characteristic speeds, establish the continuous mechanical-energy balance with geometric work and frictional dissipation, and obtain a weak entropy inequality for vanishing-viscosity limits under stated convergence assumptions. On the rotating sphere with a fixed round metric, the skew-adjoint force represents Coriolis effects. We construct a kinetic finite-volume scheme combining local fluxes, hydrostatic reconstruction, and geometric source terms. A second-order time integrator uses independently restricted steps at its two Euler stages. The scheme conserves discrete water volume, preserves nonnegative depth under a sufficient CFL condition, and is well balanced for lakes at rest, including wet/dry interfaces. Six-hour simulations of a localized surface disturbance over Earth topography with Manning bottom friction maintain positive depths and conserve water volume to near machine precision. Comparison with a spherical central-upwind method shows similar surface-wave propagation and gauge time series, with remaining differences in late-time local extrema.
\end{abstract}

\section{Introduction}

A localized disturbance of the ocean surface can propagate over distances at which the curvature of the Earth and planetary rotation become essential. Shallow-water models describe this motion through the evolution of layer depth and horizontal momentum, while bottom topography controls both wave propagation and equilibrium states. For a standard treatment of the shallow-water equations within the theory and numerical approximation of hyperbolic conservation laws, see LeVeque~\cite{leveque2002finite}. A fundamental contribution to the numerical approximation of hyperbolic conservation laws was made by Roe \cite{Roe1981}, who introduced an approximate Riemann solver based on a conservative linearization of the flux Jacobian. By constructing a matrix that satisfies the Rankine--Hugoniot jump relation and possesses a complete set of real eigenvalues and eigenvectors, Roe established a framework for resolving discontinuities without solving the full nonlinear Riemann problem. This approach has had a profound influence on the development of numerical methods for hyperbolic systems, including the shallow-water equations. Roe also contributed directly to the numerical treatment of shallow-water and related free-surface flows. Kast, Fidkowski, and Roe \cite{KastFidkowskiRoe2011} developed an entropy-adjoint approach for adaptive solutions of the shallow-water equations, while Vasconcelos, Wright, and Roe \cite{VasconcelosWrightRoe2006a,VasconcelosWrightRoe2006b,VasconcelosWrightRoe2009} investigated numerical methods for transitions between free-surface and pressurized flows, including the treatment of hydraulic bores and numerical oscillations. Together, these contributions highlight the importance of the hyperbolic structure, conservation properties, and accurate resolution of discontinuities in computational fluid dynamics. A useful formulation must therefore connect geometry, conservation, and force balance. Its numerical approximation must also preserve nonnegative water depth and avoid generating waves from a motionless free surface. This well-balanced property, studied by Berm\'udez and V\'azquez and by Garc\'ia-Navarro and V\'azquez-Cend\'on~\cite{bermudez1994upwind,garcia2000source}, is particularly important in tsunami modelling, where small surface disturbances propagate over strongly varying bathymetry and may be tiny compared with the ocean depth. If the discrete pressure and topographic forces fail to balance at rest, their numerical residual can generate spurious currents and waves that contaminate the tsunami signal. Preserving the resting state allows the method to resolve perturbations about that equilibrium without introducing this artificial forcing. These requirements motivate the present study of a kinetic representation of shallow-water equations on Riemannian manifolds and its realization as a finite-volume scheme on the rotating sphere.

The shallow-water approximation nevertheless has a physical limitation: the water depth must be small relative to the horizontal scales of the waves of interest. In the deep ocean, shorter waves can violate this assumption, and frequency dispersion neglected by the hydrostatic model can become important, particularly over long propagation distances. In that regime, a dispersive model, such as an appropriate Boussinesq-type system, should be used; Kirby et al.\ \cite{kirby2013dispersive} develop such equations on a rotating sphere and examine sensitivity to source width, dispersion, and Coriolis effects. Conventional Boussinesq approximations also have a limited range of validity, so they are not a universal replacement for arbitrary deep-water waves. Our use of shallow-water equations over the entire sphere is a leading-order long-wave approximation. In our simulations, the prescribed Gaussian disturbance has a characteristic horizontal width of approximately $222\ \mathrm{km}$, much larger than the kilometre-scale ocean depths in the experiment. This scale separation motivates a common hydrostatic model for studying global propagation and the geometric properties of the scheme. 

An intrinsic description separates the physical equations from the coordinates used to express them. The metric determines lengths, momentum norms, and volume, while covariant differentiation accounts for changes of tangent frames. Spherical-coordinate equations incorporating these effects are equivalent to the intrinsic equations; replacing Cartesian derivatives by angular derivatives alone is insufficient. Working on a general manifold provides a common framework for different reference surfaces. Rotational flattening motivates oblate spheroids, as in the nonspherical shallow-water models of Staniforth and White \cite{staniforth2015}, which also include latitude-dependent gravity. For planets and satellites subject to tidal deformation, triaxial ellipsoids may be more appropriate: a companion's stronger gravitational pull on the near side tends to stretch the body toward it, potentially producing three unequal principal axes when combined with rotation. The spatially varying curvature and metric factors of these surfaces cannot be captured by changing a single radius in a round-sphere formulation. The intrinsic approach incorporates this geometry through the metric while retaining the same kinetic moment construction and covariant balance laws. For planets or moons that support a suitable liquid layer, these geometries provide a motivation for extending the present framework beyond the sphere, without presuming the existence of an ocean on any particular body. Terrain-following flows motivate the arbitrary-topography models of Bouchut and Westdickenberg \cite{bouchut2004topography}. Such applications also clarify the scope of a geometric formulation: a metric specifies the geometry, but the effective gravity and other forces must be chosen consistently with the physical depth-averaging assumptions.

Time-dependent geometry is motivated, for example, by Jupiter's moon Europa, whose ice shell deforms under Jupiter's tidal forcing during its orbit. Models of this response couple the shell and subsurface ocean; see Beuthe~\cite{beuthe2015tides}. A deforming reference surface induces a time-dependent metric on a fixed parameter manifold, changing both lengths and area elements. This motivates allowing a prescribed family $g(t)$ in the intrinsic theory. The resulting model accounts for changing geometric volume and the associated mechanical work. It is not a complete Europa ocean model: tidal potentials, ice-shell stresses, and the kinematics and forces of an embedded moving surface require a separate physical derivation. The spherical specialization and all numerical experiments below retain a fixed metric.

Covariant shallow-water models and numerical methods on curved surfaces are well established. M\"uller \cite{mueller1992} develops a rotating-sphere formulation and examines its angular-momentum budget. Arpaia et al.\ \cite{arpaia2022} combine ambient projection with intrinsic tangential geometry in a well-balanced discontinuous Galerkin method, with applications to tsunamis and storm surges. Montoya, Rueda-Ram\'irez, and Gassner \cite{montoya2026} construct covariant discontinuous spectral-element methods that preserve mass and still-water states and conserve or dissipate discrete energy. Gonz\'alez, Castro, and Mac\'ias \cite{gonzalez2026} address geostrophic equilibria through asymptotically well-balanced finite-volume reconstructions in spherical coordinates. These works provide the geometric and numerical context for the present approach, whose emphasis is the connection between intrinsic kinetic moments and a spherical kinetic discretization.

The kinetic starting point is the Saint-Venant formulation of Perthame and Simeoni \cite{perthame2001}. Macroscopic depth and momentum are recovered as moments of a nonnegative distribution. Their scheme preserves nonnegative depth and still-water equilibria and satisfies a discrete entropy inequality. Related constructions include the kinetic avalanche model of Mangeney-Castelnau et al.\ \cite{mangeney2003} and the multilayer Saint-Venant system with mass exchanges of Audusse et al.\ \cite{audusse2011}. Kinetic transport on manifolds also has precedents: Love and Cianci \cite{love2012} derive fluid equations from a surface Boltzmann equation through a Chapman--Enskog expansion. Their Maxwell--Boltzmann closure differs from the hydrostatic shallow-water closure considered here.

For numerical treatment of topography, hydrostatic reconstruction was introduced by Audusse et al.\ \cite{audusse2004hydrostatic}. Audusse and Bristeau \cite{audusse2005} combine it with a two-dimensional kinetic solver and a positivity-preserving, well-balanced second-order reconstruction on unstructured meshes. These are direct precedents for our spherical discretization. Entropy requires a separate analysis: Audusse et al.\ \cite{audusse2016} establish a fully discrete kinetic entropy inequality for hydrostatic reconstruction with an error term that vanishes under mesh refinement. A complementary approach is the well-balanced, positivity-preserving central-upwind method of Kurganov and Petrova \cite{kurganovPetrova2007}, which provides the basis for the spherical comparison method used in our experiments.

We first formulate the kinetic closure on the cotangent bundle $T^*\Omega$ of a fixed smooth manifold $\Omega$ with a prescribed smooth family of Riemannian metrics $g(t)$. The canonical Poisson bracket describes Hamiltonian transport, and the metric defines intrinsic integration over momentum fibers. Mass, momentum, and tensor second-moment conditions yield the shallow-water balance laws with topography, a skew-adjoint force including spherical Coriolis effects, and bottom friction represented by a nonnegative scalar damping rate. An explicit compactly supported Gibbs equilibrium realizes these conditions. The derivation distinguishes scalar and tensor second moments and identifies their dimension-dependent relationship. We also obtain the conservative-variable quasilinear formulation, characteristic wave speeds, mass conservation with the evolving volume element, fixed-metric rest-state properties, and the continuous mechanical-energy balance with geometric work and frictional dissipation. Under the stated convergence assumptions, a viscous regularization yields the corresponding entropy inequality in the weak sense. This is a moment closure; a kinetic relaxation limit is not proved.

We then construct a spherical kinetic finite-volume method using local normal fluxes, hydrostatic reconstruction, and consistent geometric, Coriolis, and bottom-friction sources. A second-order time integrator allows separate CFL-restricted steps at its two Euler stages. We prove conservation of discrete water volume, preservation of nonnegative depth under a sufficient CFL condition, and preservation of lakes at rest, including wet/dry interfaces. The intrinsic friction rate includes Manning drag as a surface-flow example, and is included in the reported experiments. Numerical experiments follow a localized disturbance over Earth topography and compare surface anomalies, physical speeds, and fixed-location time series with a spherical central-upwind implementation. The experiments illustrate basin-scale propagation and agreement between the methods; they do not establish a convergence rate or validate coastal inundation.

Section~2 introduces the geometric framework. Section~3 derives the intrinsic equations and their main properties, and Section~4 specializes them to the sphere. Section~5 presents the kinetic finite-volume scheme and its conservation, positivity, and well-balancing results. Section~6 reports the numerical experiments, and Section~\ref{sec:conclusions} presents the conclusions and directions for further work. Appendix~\ref{app:semiclassical} discusses an alternative approximation based on semiclassical quantization.


\section{The ambient space}\label{sec:geometry}

The geometric material in this section is standard and is recalled to make the presentation self-contained and fix notation. For a detailed treatment of manifolds, tensor calculus, differential forms, and their applications to mechanics, see Abraham, Marsden, and Ratiu~\cite{abraham1988manifolds}. Readers familiar with Riemannian geometry and the canonical symplectic structure of the cotangent bundle may proceed directly to Section~3 and consult this section as needed.

We formulate an intrinsic barotropic model on Riemannian manifolds of arbitrary dimension, with the usual surface shallow-water interpretation on the sphere when $n=2$, or the classical shallow water equations on the plane. The metric determines the geometry entering transport and force balance, independently of any embedding. This section introduces the geometric objects needed for a formulation that is consistent under changes of local coordinates. 

Let $\Omega$ be a fixed smooth manifold of dimension $n\geq1$, equipped with a prescribed smooth family of positive-definite Riemannian metrics $g(t)$, $t\in[0,T]$. No embedding, orientation, constant curvature, or compactness is required for the local derivation. We work in the interior if $\Omega$ has a boundary; global conservation then requires boundary conditions in those cases. We write $g_x=g(t,x)$, suppressing time in point subscripts and in the objets $\nabla$, $\sharp$, $\flat$, $dV_g$, and $d\mu_x$ (to be defined below); these objects always use the metric at the current time. Time derivatives are taken at fixed base and cotangent coordinates on this fixed manifold, and coordinate changes are independent of time. Thus this formulation does not prescribe an embedding or the motion of a coordinate frame. 

At each $x\in\Omega$, the tangent space $T_x\Omega$ consists of tangent vectors, and the cotangent space $T_x^*\Omega$ is its dual, consisting of linear functionals on tangent vectors. The metric is an inner product
\[
g_x:T_x\Omega\times T_x\Omega\longrightarrow\mathbb R.
\]
In local coordinates, for $v=v^i\partial/\partial x^i$ and $w=w^j\partial/\partial x^j$, it is given by
\[
g_x(v,w)=(g_x)_{ij}v^iw^j,\qquad
(g_x)_{ij}=g_x(\partial/\partial x^i,\partial/\partial x^j).
\]
Here and below, repeated upper and lower indices are summed from $1$ to $n$. Positive definiteness makes the matrix $((g_x)_{ij})$ invertible. Its inverse $((g_x)^{ij})$ defines the \emph{dual metric}, also called the \emph{cometric}, on the cotangent space:
\[
g_x^{-1}:T_x^*\Omega\times T_x^*\Omega\longrightarrow\mathbb R,
\qquad g_x^{-1}(\xi,\eta)=(g_x)^{ij}\xi_i\eta_j,
\]
where $\xi=\xi_i\,dx^i$ and $\eta=\eta_i\,dx^i$. Thus $g_x$ measures lengths and angles of tangent vectors, whereas $g_x^{-1}$ does so for covectors. In particular,
\[
|v|_{g_x}^2=g_x(v,v),\qquad
|\xi|_{g_x^{-1}}^2=g_x^{-1}(\xi,\xi).
\]
Both inner products are intrinsic; the inverse-matrix formula expresses their relationship in any local coordinate basis. The superscript $-1$ denotes this inverse metric, not the reciprocal of a scalar.

\begin{note}
The constant $g_{\mathrm{grav}}>0$ used below denotes gravitational acceleration; the subscript distinguishes this scalar from the Riemannian metric $g$.
\end{note}

\subsection{The cotangent bundle and its symplectic form}

\begin{figure}[htp]
\centering
\includegraphics[width=.6\textwidth]{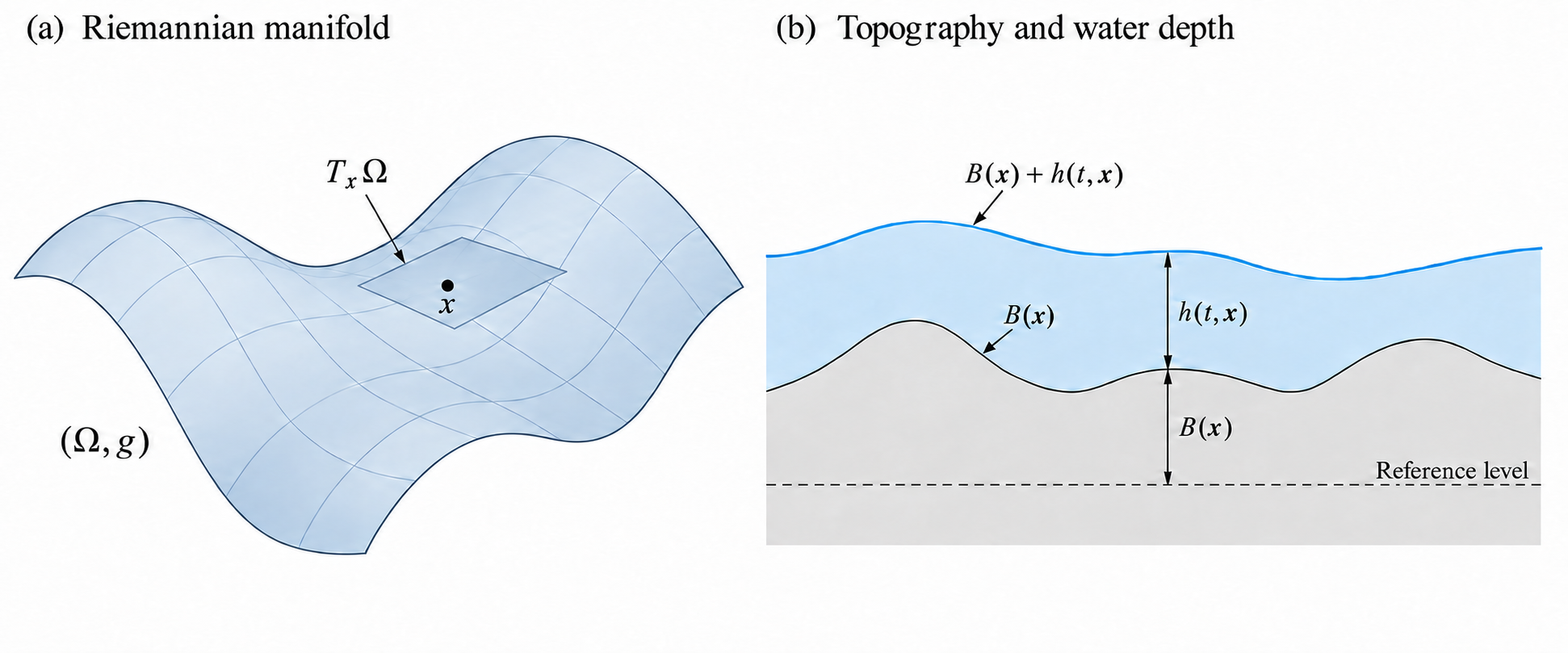}
\caption{Left panel: Schematic of the ambient space for the shallow water equations. Right panel: Description of the involved variables.}
\label{fig:riemannian-schematic}
\end{figure}

Let $T^*\Omega$ denote the cotangent bundle, and $\pi:T^*\Omega\to \Omega$ denote the cotangent-bundle projection. A point of $T^*\Omega$ is a pair $(x,\xi)$, where $x\in \Omega$ and $\xi\in T_x^*\Omega$. The phase space $T^*\Omega$ has the structure of a smooth $2n$-dimensional manifold. 

The shallow-water equations are written in terms of the conservative variables water depth $h(t,x)$ and momentum $m(t,x)=h(t,x)u(t,x)$, for $x\in\Omega$ and $t\in[0,T]$. The depth satisfies mass conservation, while momentum evolves according to Newton's second law, accounting for gravity, hydrostatic pressure, and dissipative bottom friction, together with Coriolis effects when rotation is included. Here, $u(t,\cdot)$ is interpreted as a covector field or, equivalently, a differential $1$-form on $\Omega$. More formally,
\begin{equation}
h: [0,T] \times \Omega \to \mathbb{R}_{\geq 0} , \; \;  u: [0,T] \times \Omega \to T^* \Omega, \; \;  \pi(u(t,x)) = x. 
\end{equation} 
That is, $u(t,x) \in T_x^*\Omega$ for each $x \in \Omega$. The left panel of Figure \ref{fig:riemannian-schematic} shows the schematic the ambient space, while the right panel exhibits the model variables and parameters. 

Our kinetic approach requires the use of the associated Poisson bracket. This bracket acts on smooth functions on this phase space:
\[
\{\cdot,\cdot\}:C^\infty(T^*\Omega)\times C^\infty(T^*\Omega)\longrightarrow C^\infty(T^*\Omega).
\]

It is well known that  the familiar canonical formula is preserved by coordinate changes on the base together with the corresponding covector transformation of the momenta. This crucial property will help us derive the model intrinsically, consistent in any local coordinate choice. We now provide more details about the Poisson bracket. 

The cotangent bundle has a distinguished one-form $\lambda$, called the tautological or Liouville one-form. At $(x,\xi)\in T^*\Omega$, it is defined by
\begin{equation}\label{eq:lambda}
\lambda_{(x,\xi)}(V)=\xi\bigl(d\pi_{(x,\xi)}V\bigr),\qquad V\in T_{(x,\xi)}(T^*\Omega).
\end{equation}
Thus $\lambda$ is a globally defined smooth one-form. We define $\omega=-d\lambda$. If $(x^1,\ldots,x^n)$ are coordinates on an open set $U\subset \Omega$, every covector over $U$ has a unique expression $\xi=\xi_i\,dx^i$, where repeated indices are summed from $1$ to $n$. The induced coordinates on $T^*U$ are $(x^1,\ldots,x^n,\xi_1,\ldots,\xi_n)$, and
\begin{equation}\label{eq:canonical}
\lambda=\xi_i\,dx^i,\qquad \omega=dx^i\wedge d\xi_i,
\end{equation}
where $\wedge$ is the wedge product.  In particular, $d\omega=0$. Let us recall the following 
\begin{definition}
The contraction (or interior product) $\iota_V\alpha$ of a differential $k$-form $\alpha$, $k\geq1$, with a tangent vector $V$ is the $(k-1)$-form obtained by inserting $V$ into its first argument:
\[
 (\iota_V\alpha)(V_1,\ldots,V_{k-1})
 =\alpha(V,V_1,\ldots,V_{k-1}),
\]
where all vectors are tangent at the same point. In particular, for the two-form $\omega$ on $T^*\Omega$, this defines the one-form
\[
 (\iota_V\omega)_{(x,\xi)}(W)
 =\omega_{(x,\xi)}(V,W),\qquad
 V,W\in T_{(x,\xi)}(T^*\Omega).
\]
\end{definition}

For a tangent vector $V=a^i\frac{\partial}{\partial x^i}+b_i\frac{\partial}{\partial \xi_i}$, the corresponding contraction is:
$
\iota_V\omega=a^i\,d\xi_i-b_i\,dx^i.
$
If this one-form vanishes, all $a^i$ and $b_i$ vanish. Hence $\omega$ is nondegenerate at every point and is a symplectic form. No Riemannian metric was needed. Nondegeneracy gives, for each smooth function $a$, a unique Hamiltonian vector field $X_a$ satisfying
\begin{equation}\label{eq:hamiltonian}
\iota_{X_a}\omega=da.
\end{equation}
\begin{definition}[Poisson bracket]
For $a,b\in C^\infty(T^*\Omega)$, let $X_a,X_b$ be their Hamiltonian vector fields with the convention~\eqref{eq:hamiltonian}. Their \emph{Poisson bracket} is the smooth function $\{a,b\}\in C^\infty(T^*\Omega)$ defined by
\begin{equation}\label{eq:bracket}
\{a,b\}=\omega(X_a,X_b)=da(X_b).
\end{equation}
\end{definition}
This definition uses only global geometric objects, so it produces a well-defined smooth function on all of $T^*\Omega$ whenever $a$ and $b$ are global smooth functions. Equations \eqref{eq:canonical} and \eqref{eq:hamiltonian} give
\begin{equation}\label{eq:localbracket}
X_a=\frac{\partial a}{\partial \xi_i}\frac{\partial}{\partial x^i}
-\frac{\partial a}{\partial x^i}\frac{\partial}{\partial \xi_i}, \; \text{ and therefore},
\{a,b\}=\sum_{i=1}^n\left(
\frac{\partial a}{\partial x^i}\frac{\partial b}{\partial \xi_i}
-\frac{\partial a}{\partial \xi_i}\frac{\partial b}{\partial x^i}\right).
\end{equation}
The Poisson bracket is bilinear, antisymmetric, and satisfies the Leibniz rule and the Jacobi identity. 


%



\subsection{The metric and intrinsic momentum integration}

For a coordinate basis $\partial/\partial x^i$ in $T_x\Omega$, let
\[
(g_x)_{ij}=g_x(\partial/\partial x^i,\partial/\partial x^j),\qquad ((g_x)^{ij})=((g_x)_{ij})^{-1},
\qquad G(t,x)=\det((g_x)_{ij}).
\]
Here $(g_x)^{ij}$ are the components of the dual metric (or cometric) $g_x^{-1}$. The metric induces the \emph{musical isomorphisms}
\[
\sharp:T_x^*\Omega\longrightarrow T_x\Omega,\qquad
\flat:T_x\Omega\longrightarrow T_x^*\Omega.
\]
The sharp map raises an index, converting a covector into a tangent vector; the flat map lowers an index and is its inverse. For $u,\xi\in T_x^*\Omega$ and $v\in T_x\Omega$, these maps and the squared dual norm are
\[
u^\sharp=(g_x)^{ij}u_j \partial/\partial x^i,\qquad
v^\flat=g_x(v,\cdot),\qquad
|\xi|_{g_x^{-1}}^2=(g_x)^{ij}\xi_i\xi_j.
\]
Thus $g_x(u^\sharp,v)=u(v)$. In particular, $(g_x)^{ij}(g_x)_{ij}=n$. 

These operations connect the covector variables of the model to physical motion. Since $u$ is a covector, the tangent vector $u^\sharp$ is the transport velocity: the mass flux is $h u^\sharp$, and mass conservation takes the form $\partial_t h+\vartheta h+\operatorname{div}_g(h u^\sharp)=0$, where $\vartheta=\partial_t\log\sqrt G$ measures the local change of geometric volume (to be derived and explained below). Conversely, the flat map expresses a physical velocity or acceleration as a covector, consistently with the momentum variable $hu$. The particle variable $\xi$ is also a covector, so its kinetic energy per unit mass is $\tfrac12|\xi|_{g_x^{-1}}^2$; the Gibbs equilibrium that will be discussed below depends on the corresponding squared distance $|\xi-u|_{g_x^{-1}}^2$ from the fluid velocity covector. 

The dual metric equips the vector space $T_x^*\Omega$ with a Euclidean measure $d\mu_x$. At fixed $(t,x)$,
\begin{equation}\label{eq:fiber-measure}
d\mu_x(\xi)=\frac{|d\xi_1\cdots d\xi_n|}{\sqrt{G(t,x)}},
\qquad dV_g(x)=\sqrt{G(t,x)}\,|dx^1\cdots dx^n|.
\end{equation}
These are positive densities and do not require an orientation. The canonical Liouville measure is
\begin{equation}\label{eq:liouville}
dL=\left|\frac{\omega^n}{n!}\right|
=|dx^1\cdots dx^n d\xi_1\cdots d\xi_n|
=dV_g(x)\,d\mu_x(\xi).
\end{equation}
The phase-space measure $dL$ is independent of time and of $g$, although both factors on the right depend on the metric. Write $\dot g=\partial_t g$ and define
\begin{equation}\label{eq:metric-volume-rate}
 \vartheta=\partial_t\log\sqrt G
 =\frac12(g_x)^{ij}\dot g_{ij},\qquad
 \partial_t dV_g=\vartheta\,dV_g,\qquad
 \partial_t d\mu_x=-\vartheta\,d\mu_x.
\end{equation}
Consequently, for a sufficiently regular integrable fiber function $a(t,x,\xi)$,
\begin{equation}\label{eq:time-fiber-integral}
 \int\partial_t a\,d\mu_x
 =\partial_t\int a\,d\mu_x+\vartheta\int a\,d\mu_x.
\end{equation}
The inverse metric also varies according to
$\partial_t(g_x)^{ij}=-(g_x)^{ia}\dot g_{ab}(g_x)^{bj}$.
These identities supply the additional terms in the time-dependent balance laws.



Divergence is intrinsic once a smooth positive density has been specified. Here \emph{density} is used in the differential-geometric sense: it is a volume element that can be integrated without choosing an orientation, thereby defining a measure. The Riemannian volume element $dV_g$ and the Liouville volume element $dL$ are examples of such geometric densities. The coefficient $\rho$ in the local expression $\nu=\rho(z)|dz^1\cdots dz^N|$ below is a coordinate weight for this volume element, not a physical mass density.

Recall that the Lie derivative measures the infinitesimal change of a geometric object along a vector field. For a smooth vector field $V$, let $\Phi_s$ denote its local flow, so that $\Phi_0$ is the identity and $\frac{d}{ds}\Phi_s(z)=V(\Phi_s(z))$. The parameter $s$ parametrizes this flow and is independent of the physical time $t$. For a smooth density $\nu$, its Lie derivative is
\[
\mathcal L_V\nu=\left.\frac{d}{ds}\right|_{s=0}\Phi_s^*\nu.
\]
Here $\Phi_s^*\nu$ is the pullback density: it measures the volume of an infinitesimal region after that region is carried along by the flow, expressed at its original location. 

\begin{definition}
Let $\nu$ be a smooth positive density and $V$ a smooth vector field on the same manifold. The divergence of $V$ with respect to $\nu$ is the unique smooth function $\operatorname{div}_{\nu}V$ satisfying
\begin{equation}\label{eq:divergence-intrinsic}
\mathcal L_V\nu=(\operatorname{div}_{\nu}V)\nu,
\end{equation}
where $\mathcal L_V$ is the Lie derivative defined above.
\end{definition}
Thus divergence measures the infinitesimal relative expansion of volume under the flow of $V$. If $z^1,\ldots,z^N$ are local coordinates, $\nu=\rho(z)|dz^1\cdots dz^N|$, and $V=V^a\partial/\partial z^a$, then
\begin{equation}\label{eq:divergence-density}
\operatorname{div}_{\nu}V
=\frac{1}{\rho}\frac{\partial}{\partial z^a}(\rho V^a).
\end{equation}
Although its expression uses coordinates and it does depend on $\rho$, the resulting scalar is independent of the chart. On $T^*\Omega$ we use the canonical Liouville density $dL$ in \eqref{eq:liouville}. In induced cotangent coordinates its density factor is one, so
\begin{equation}\label{eq:phase-divergence}
V=a^i\frac{\partial}{\partial x^i}+b_i\frac{\partial}{\partial\xi_i}
\quad\Longrightarrow\quad
\operatorname{div}_{dL}V
=\frac{\partial a^i}{\partial x^i}+\frac{\partial b_i}{\partial\xi_i}.
\end{equation}
In arbitrary phase-space coordinates one must instead retain the density factor in \eqref{eq:divergence-density}. 


A vector field is called vertical if $d\pi(V)=0$: its motion stays within a fiber $T_x^*\Omega$, leaving the base point $x$ fixed. Such a field has the form $V=F_i(x,\xi)\partial/\partial\xi_i$, and
\begin{equation}\label{eq:vertical-divergence}
\operatorname{div}_{dL}V=\operatorname{div}_{\xi}F
=\frac{\partial F_i}{\partial\xi_i}.
\end{equation}
Here $\operatorname{div}_{\xi}$ is also the divergence relative to the fiber measure $d\mu_x$: its factor $G(t,x)^{-1/2}$ is constant when differentiating in $\xi$. For comparison, the divergence of a vector field $v=v^i\partial/\partial x^i$ on the base relative to its Riemannian density is
\[
\operatorname{div}_g v
=\frac{1}{\sqrt{G}}\frac{\partial}{\partial x^i}(\sqrt{G}\,v^i).
\]
The different coordinate expressions on the base and on phase space reflect their respective densities; both define intrinsic scalars, including when $\Omega$ is nonorientable.


\section{Derivation of the model using a kinetic approach}

\subsection{Background: one-dimensional shallow-water equations}

The kinetic approach represents the macroscopic fluid variables as moments of a nonnegative distribution in an auxiliary momentum variable at a microscopic level. For the shallow-water equations, the central example is the construction of Perthame and Simeoni \cite{perthame2001} for the one-dimensional Saint-Venant system with bottom topography. Their kinetic formulation provides the basis for a numerical scheme that preserves nonnegative water depth and still-water equilibria, satisfies an entropy inequality, and accommodates discontinuous bottom profiles. 

To illustrate the moment mechanism and prepare for the general formulation, we first neglect bottom friction and Coriolis effects. Equip the base with the standard fixed metric and consider a flat one-dimensional domain with coordinate $x_1\in[a,b]\subset\mathbb R$, scalar fluid velocity $u_1(t,x_1)$, and kinetic variable $\xi_1\in\mathbb R$. The subscript $1$ distinguishes the variables in this one-dimensional setting from those used in the general formulation. Although the interval has a boundary, we derive the evolution equations only in its interior. The same analysis applies to the unit circle $S^1$, using a periodic arclength coordinate.

At a microscopic level, the particle interactions and collisions are modeled statistically through a density of particles denoted by $M_1(t,x_1,\xi_1)$. The connection with the macroscopic level is through the following relations
\begin{equation}\label{eq:flat-kinetic-motivation}
\int_{\mathbb R} M_1 \,d \xi_1=h,\qquad
\int_{\mathbb R} \xi_1 M_1 \,d \xi_1 =h u_1,\qquad
\int_{\mathbb R} \xi_1^2 M_1\,d \xi_1 = h u_1^2+\frac12g_{\mathrm{grav}}h^2.
\end{equation}
Here $g_{\mathrm{grav}}$ is the gravitational acceleration. For $h>0$, these relations hold for the Gibbs equilibrium
\[
M_1(t,x_1,\xi_1)= \sqrt{g_{\mathrm{grav}} h(t,x_1)} \frac{\sqrt{2}}{\pi g_{\mathrm{grav}}} \sqrt{\max\left( 1-\frac{1}{2}\left( \frac{\xi_1-u_1(t,x_1)}{\sqrt{g_{\mathrm{grav}} h(t,x_1)}}  \right)^2,0 \right)}.
\]

Thus the second moment considers a hydrostatic pressure. With $B(x_1)$ denoting bottom topography, the corresponding kinetic transport equation has the form
\begin{equation}
\label{eq:KineticEqnDm1}
\partial_t M_1+\xi_1 \,\partial_{x_1}  M_1 -g_{\mathrm{grav}} \partial_{x_1} B \,\partial_{\xi_1} M_1 = C_1,
\qquad
\int_{\mathbb R} C_1 \,d \xi_1 =\int_{\mathbb R} \xi_1 C_1\,d \xi_1=0.
\end{equation}
Here $C_1$ is a collision term with vanishing mass and momentum moments.  Integration against $1$ and $\xi_1$, with sufficient decay in $\xi_1$, provides the regular frictionless shallow water equations
\[
\partial_t h+\partial_{x_1}(h u_1 )=0,\qquad
\partial_t(h u_1)+\partial_{x_1} \!\left(hu_1^2+\frac12g_{\mathrm{grav}}h^2\right)
=-g_{\mathrm{grav}} \; h\,\partial_{x_1} B.
\]

A key structural observation is that equation \eqref{eq:KineticEqnDm1} is a Hamiltonian transport equation with a collision term. It can be written as
\begin{equation}
\label{eq:flat-kinetic-bracket}
\partial_t M_1 +\left\{ M_1(t,x_1,\xi_1), H(x_1,\xi_1)  \right\} =C_1(t,x_1,\xi_1),
\end{equation}
where $\left\{ \cdot, \cdot \right\}$ is the Poisson bracket  \eqref{eq:localbracket}, and $H(x_1,\xi_1)=\frac{\xi_1^2}{2}+g_{\mathrm{grav}}B(x_1)$ is the associated Hamiltonian. The significance of this expression is that the Poisson bracket is intrinsically defined on the cotangent bundle of any smooth manifold, through its canonical symplectic form, as recalled in Section~\ref{sec:geometry} and definition~\eqref{eq:bracket}. Its definition requires neither a Riemannian metric nor an embedding. The metric enters through the kinetic energy in $H$, while the bracket provides the same coordinate-independent transport mechanism on every cotangent bundle. This observation is the basis for the extension to Riemannian manifolds in the next subsection.

The Hamiltonian formulation also connects the model to semiclassical analysis. Phase-space observables can be quantized as operators, and, for smooth symbols in the appropriate calculus, the Poisson transport term $\{M,H\}$ is the principal symbol of the scaled commutator $(i/\hbar)[H^W,M^W]$, where $H^W$ and $M^W$ denote Weyl quantizations and $\hbar$ is the semiclassical parameter \cite{zworski}. Thus the same structure supports both an intrinsic classical formulation and a possible operator approximation. This correspondence is semiclassical, rather than an exact identity at finite $\hbar$; the nonsmooth equilibrium requires additional care. Appendix~\ref{app:semiclassical} discusses this direction and its limitations.

\subsection{Generalization to Riemannian manifolds}

The relations in equations \eqref{eq:flat-kinetic-motivation} and \eqref{eq:KineticEqnDm1}  explain the underlying mechanism in our variables: transport supplies the moment fluxes, the second moment supplies pressure, and the force in momentum space supplies the topographic source.

Our construction retains this moment mechanism on $T^*\Omega$ in the general case of a Riemannian manifold. Let $M(t,x,\xi)\geq0$ be a scalar distribution relative to $dL$, interpreted as the density of particles. Its spatial density is
\begin{equation}\label{eq:particle-density}
h(t,x)=\int_{T_x^*\Omega}M(t,x,\xi)\,d\mu_x(\xi)
=\frac1{\sqrt{G(t,x)}}\int_{\mathbb R^n}M(t,x,\xi)\,d\xi_1\cdots d\xi_n.
\end{equation}




Let $u(t,x)\in T_x^*\Omega$ be the mean particle covector, representing fluid velocity, for $h>0$. We require the following generalization of the moment conditions used previously:
\begin{equation}\label{eq:equilibrium-targets}
\begin{aligned}
\int_{T_x^*\Omega}M\,d\mu_x&=h, \; \; \; 
\int_{T_x^*\Omega}\xi M\,d\mu_x=hu,\\
\int_{T_x^*\Omega}|\xi|_{g_x^{-1}}^2M\,d\mu_x
&=h|u|_{g_x^{-1}}^2+\frac n2 g_{\mathrm{grav}}h^2,\\
\int_{T_x^*\Omega}\xi\otimes\xi M\,d\mu_x
&=h\,u\otimes u+\frac12g_{\mathrm{grav}}h^2g_x.
\end{aligned}
\end{equation}
The tensor condition prescribes the pressure $p(h)=g_{\mathrm{grav}}h^2/2$. Contracting the tensor identity with the inverse metric gives the scalar condition, since $(g_x)^{ij}(g_x)_{ij}=n$. On the other hand, for the case $n=1$, the last two equations are equivalent, as it is done in equation \eqref{eq:flat-kinetic-motivation}.

\begin{proposition}
Let $\Omega$ be an $n$-dimensional manifold, $n\geq1$, with a smooth family of Riemannian metrics $g(t)$, and let $g_{\mathrm{grav}}>0$. For a depth $h(t,x)>0$ and a covector $u(t,x)\in T_x^*\Omega$, define
\begin{equation}\label{eq:general-gibbs-equilibrium}
M(t,x,\xi)=A_n(h(t,x))
\sqrt{\max\left(1-
\frac{2|\xi-u(t,x)|_{g_x^{-1}}^2}{(n+3)g_{\mathrm{grav}}h(t,x)},0\right)}.
\end{equation}
where
\begin{equation}\label{eq:general-equilibrium-parameters}
A_n(h)=\frac{2\,\Gamma((n+3)/2)}{\pi^{(n+1)/2}}
\left(\frac{2}{(n+3)g_{\mathrm{grav}}}\right)^{n/2}h^{1-n/2}.
\end{equation} Here $\Gamma$ is the Euler gamma function. At points where $h=0$, set $M=0$. Then $M$ is nonnegative, is independent of the choice of coordinates, and satisfies all the moment conditions in \eqref{eq:equilibrium-targets}. For $h>0$, its momentum support is the intrinsic closed ball
\[
\left\{\xi\in T_x^*\Omega:
|\xi-u(t,x)|_{g_x^{-1}}\leq
\sqrt{\frac{n+3}{2}g_{\mathrm{grav}}h(t,x)}\right\}.
\]
\end{proposition}
\begin{proof}
Fix $(t,x)$ with $h>0$ and use the metric $g(t,x)$. No time differentiation is involved in these moment identities. We seek a radial equilibrium with unknown amplitude $A>0$ and support radius $a>0$, to be determined from the moment conditions in \eqref{eq:equilibrium-targets}. Let us consider the profile
\begin{equation}\label{eq:general-square-root-profile}
M(t,x,\xi)=A\sqrt{\max\left(1-
\frac{|\xi-u(t,x)|_{g_x^{-1}}^2}{a^2},0\right)}.
\end{equation}
This construction uses only the cotangent vector $\xi-u$ and the dual metric $g_x^{-1}$, so it is independent of coordinates or an orthonormal basis, provided $A$ and $a$ are scalars.

Translation by $u$ preserves $d\mu_x$. Choose a $g_x^{-1}$-orthonormal basis of $T_x^*\Omega$, write $\eta=\xi-u$, and use Euclidean polar coordinates $r=|\eta|_{g_x^{-1}}$. With $|S^{n-1}|=2\pi^{n/2}/\Gamma(n/2)$, direct integration gives
\begin{align}
\int_{T_x^*\Omega} M\,d\mu_x
&=A|S^{n-1}|\int_0^a r^{n-1}\sqrt{1-r^2/a^2}\,dr
=\frac{\pi^{(n+1)/2}A a^n}{2\,\Gamma((n+3)/2)},\label{eq:general-equilibrium-mass}\\
\int_{T_x^*\Omega}|\eta|_{g_x^{-1}}^2M\,d\mu_x
&=\frac{n\,\pi^{(n+1)/2}A a^{n+2}}{2(n+3)\,\Gamma((n+3)/2)},\label{eq:general-equilibrium-centered}
\end{align}
The mass condition therefore determines
$
A=\frac{2\,\Gamma((n+3)/2)}{\pi^{(n+1)/2}}\frac{h}{a^n}.
$ 
Radial symmetry gives $\int\eta M\,d\mu_x=0$, so the momentum condition $\int\xi M\,d\mu_x=hu$ is then automatic. Using the mass normalization, we obtain
\begin{equation}\label{eq:general-centered-tensor}
\int_{T_x^*\Omega}\eta\otimes\eta\,M\,d\mu_x
=\frac{h a^2}{n+3}\,g_x.
\end{equation}
Expanding $\xi\otimes\xi=(u+\eta)\otimes(u+\eta)$ and using the vanishing centered first moment, the tensor condition in \eqref{eq:equilibrium-targets} therefore requires
\[
\frac{h a^2}{n+3}=\frac12g_{\mathrm{grav}}h^2.
\]
Since $h>0$ and $a>0$, this determines
\[
a^2=\frac{n+3}{2}g_{\mathrm{grav}}h,
\qquad
A=\frac{2\,\Gamma((n+3)/2)}{\pi^{(n+1)/2}}\left(\frac{2}{(n+3)g_{\mathrm{grav}}}\right)^{n/2}h^{1-n/2}
=A_n(h).
\]
Substitution into \eqref{eq:general-square-root-profile} gives the stated equilibrium. Taking the metric trace of the tensor moment identity gives the scalar second-moment condition, since $(g_x)^{ij}(g_x)_{ij}=n$. Nonnegativity and the stated support follow from the profile. At $h=0$, every moment identity holds by setting $M=0$, independently of $u$.
\end{proof}
For $n=1$ it reduces to the one-dimensional Gibbs equilibrium above, while for $n=2$ (like in the case of the sphere below) the prefactor is independent of $h$.


Let $B\in C^\infty(\Omega)$ be a prescribed, time-independent potential topography. Define
\begin{equation}\label{eq:sw-hamiltonian}
H:[0,T]\times T^*\Omega\longrightarrow\mathbb R,\qquad H(t,x,\xi)=\frac12g_x^{-1}(\xi,\xi)+g_{\mathrm{grav}}B(x).
\end{equation}
This is a global smooth time-dependent Hamiltonian for every smooth metric family. The potential term is understood as pulled back by $\pi$. The canonical bracket does not depend on the metric, but this Hamiltonian does.

A general manifold has no distinguished cross product or Coriolis parameter. To allow a rotation-type force, prescribe a smooth family of bundle endomorphisms $\mathcal A_x=\mathcal A(t,x):T_x^*\Omega\to T_x^*\Omega$ that is skew-adjoint with respect to the current metric:
\[
g_x^{-1}(\mathcal A_x\alpha,\beta)
=-g_x^{-1}(\alpha,\mathcal A_x\beta).
\]
The choice $\mathcal A=0$ is always available and requires no additional geometric structure. On an oriented two-dimensional Riemannian manifold, let $J_x$ denote rotation by $+\pi/2$ in the tangent plane and define its action on covectors by $\mathcal J_x\xi=(J_x\xi^\sharp)^\flat$. The choice $\mathcal A_x=f(x)\mathcal J_x$ then represents the Coriolis force when $f$ is the prescribed Coriolis parameter.

In order to include bottom friction, prescribe a nonnegative scalar rate
$\gamma=\gamma(x,h,|u|_{g_x^{-1}})$, with units of inverse time, and set
\begin{equation}\label{eq:friction-force}
 F(t,x,\xi)=-\mathcal A_x\xi-\gamma u.
\end{equation}
The rate is assumed finite and sufficiently regular on wet states $h>0$;
at a dry state $h=0$, $hu=0$, we set $u=0$ and $\gamma=0$. The bottom-friction force is constant across each momentum fiber, although
it depends on the moments of $M$. Thus
$\operatorname{div}_{\xi}F=-\operatorname{tr}\mathcal A_x=0$.
Unlike the rotation term, friction does work:
$\int g_x^{-1}(\xi,-\gamma u)M\,d\mu_x=-\gamma h|u|_{g_x^{-1}}^2$.
Using $-\gamma u$ rather than $-\gamma\xi$ avoids additional damping of
microscopic fluctuations around the mean velocity.

For two-dimensional shallow-water flows, an empirical example is Manning
friction~\cite{geoclawManning},
\begin{equation}\label{eq:manning-friction}
 \gamma=\frac{g_{\mathrm{grav}}n_{\mathrm M}(x)^2}{h^{4/3}}
              |u|_{g_x^{-1}},\qquad h>0,
\end{equation}
where $n_{\mathrm M}\geq0$ is the roughness coefficient, with units
$\mathrm{s\,m^{-1/3}}$, distinct from the manifold dimension $n$.
The general-dimensional theory uses a prescribed scalar $\gamma\geq0$;
the empirical interpretation of Manning's law concerns surface flows.
Its singular dependence on depth requires care near dry states.
Setting $\gamma=0$ recovers the frictionless model.
We assume that the microscopic distribution satisfies
\begin{equation}\label{eq:sw-kinetic}
\frac{\partial M}{\partial t}+\{M,H\}
+\operatorname{div}_{\xi}(FM)=C,
\qquad
\int C\,d\mu_x=0,\quad \int\xi C\,d\mu_x=0.
\end{equation}
The fiber divergence is computed with $(t,x)$ fixed. Since $dL$ is fixed, the kinetic equation needs no additional volume term; the time dependence of $H$ enters the Hamiltonian vector field at each time. This extra force is separate from the canonical Hamiltonian dynamics and from the momentum-conserving collision term.



Write $\xi^j=(g_x)^{jk}\xi_k$, $m_i=hu_i$, and
$m^j=(g_x)^{jk}m_k$. Temporarily, let
$
K_i{}^j=\int\xi_i\xi^j M\,d\mu_x.
$
This is the mixed second-moment tensor, not an additional closure variable. Multiplying the conservative kinetic equation by $1$ and $\xi_i$ and integrating with respect to the momentum Lebesgue measure gives
\begin{align}
\frac{\partial h}{\partial t}+\vartheta h
+\frac1{\sqrt G}\frac{\partial}{\partial x^j}(\sqrt G\,m^j)&=0,
\label{eq:local-mass}\\
\frac{\partial m_i}{\partial t}+\vartheta m_i
+\frac1{\sqrt G}\frac{\partial}{\partial x^j}(\sqrt G\,K_i{}^j)
&=-\frac12\frac{\partial(g_x)^{ab}}{\partial x^i}
\int\xi_a\xi_bM\,d\mu_x
-g_{\mathrm{grav}}h\frac{\partial B}{\partial x^i}
-(\mathcal A_xm)_i-\gamma m_i.
\label{eq:local-momentum}
\end{align}
The factors $\sqrt G$ appear because momentum Lebesgue integration equals $\sqrt G$ times intrinsic fiber integration. The terms $\vartheta h$ and $\vartheta m_i$ follow from differentiating this time-dependent factor, as in \eqref{eq:time-fiber-integral}. At each time, let $\nabla$ be the Levi--Civita connection determined by $g(t)$. Its Christoffel symbols $\Gamma^k_{ij}$ are the coordinate coefficients defined by
\[
\nabla_{\partial_i}\partial_j=\Gamma^k_{ij}\partial_k,
\qquad \partial_i=\frac{\partial}{\partial x^i}.
\]
Thus $\nabla$ denotes intrinsic covariant differentiation, while $\Gamma^k_{ij}$ accounts for the variation of the coordinate basis. For a vector field $v=v^j\partial_j$ and a covector field $u=u_j\,dx^j$, respectively,
\[
\nabla_i v^j=\partial_i v^j+\Gamma^j_{ik}v^k,
\qquad
\nabla_i u_j=\partial_i u_j-\Gamma^k_{ij}u_k.
\]
The opposite signs reflect the duality between vectors and covectors. These coefficients depend on the coordinates and can be nonzero even for a flat metric expressed in curvilinear coordinates. For the Levi--Civita connection, they are given by
\[
\Gamma^k_{ij}=\frac12(g_x)^{k\ell}
\left(\frac{\partial(g_x)_{j\ell}}{\partial x^i}
+\frac{\partial(g_x)_{i\ell}}{\partial x^j}
-\frac{\partial(g_x)_{ij}}{\partial x^\ell}\right).
\]
Metric compatibility and symmetry of the second moment imply
\[
-\frac12\frac{\partial(g_x)^{ab}}{\partial x^i}
\int\xi_a\xi_b M\,d\mu_x
=\Gamma^k_{ji}K_k{}^j.
\]
For a mixed tensor,
\[
\nabla_j K_i{}^j
=\frac1{\sqrt G}\frac{\partial}{\partial x^j}(\sqrt G\,K_i{}^j)
-\Gamma^k_{ji}K_k{}^j.
\]
Here $\nabla_j$ denotes covariant differentiation along $\partial/\partial x^j$, not partial differentiation of components. Thus the metric-derivative term in \eqref{eq:local-momentum} is exactly the connection term required for an intrinsic covector divergence. This calculation verifies the result for arbitrary smooth $g$, without flatness or constant-curvature assumptions.


The tensor moment in \eqref{eq:equilibrium-targets} gives
\[
K_i{}^j=hu_i u^j+p(h)\delta_i{}^j,
\qquad p(h)=\frac12g_{\mathrm{grav}}h^2.
\]
Define the divergence of a covariant two-tensor $T$ by
$(\operatorname{div}_g T)_i=(g_x)^{jk}\nabla_j T_{ik}$.
Metric compatibility implies $\operatorname{div}_g(p g)=dp$. We summarize the resulting moment closure in the following theorem.

\begin{theorem}
\label{thm:intrinsic-shallow-water}
Let $\Omega$ be a fixed smooth $n$-dimensional manifold with a prescribed smooth Riemannian metric $g(t)$ and volume rate $\vartheta$ from \eqref{eq:metric-volume-rate}. Let $g_{\mathrm{grav}}>0$, and let $B$ be smooth and independent of time. Let $\mathcal A_x=\mathcal A(t,x):T_x^*\Omega\to T_x^*\Omega$ be a prescribed smooth family of endomorphisms, skew-adjoint with respect to $g(t)^{-1}$. Suppose that a nonnegative distribution $M$ satisfies the kinetic equation~\eqref{eq:sw-kinetic}, with Hamiltonian~\eqref{eq:sw-hamiltonian}, force $F=-\mathcal A_x\xi-\gamma u$, and collision term satisfying
\[
 \int_{T_x^*\Omega}C\,d\mu_x=0,\qquad
 \int_{T_x^*\Omega}\xi C\,d\mu_x=0.
\]
Assume the moment closure~\eqref{eq:equilibrium-targets} and sufficient regularity and fiber decay to differentiate the moments and integrate by parts without boundary terms in momentum space. Then the depth $h$ and covector momentum $hu$ satisfy
\begin{equation}\label{eq:sw-forward}
\begin{aligned}
\frac{\partial h}{\partial t}+\vartheta h + \operatorname{div}_g(hu^\sharp)
&= 0,\\
\frac{\partial(hu)}{\partial t}+\vartheta hu
+\operatorname{div}_g(hu\otimes u)
+\frac{g_{\mathrm{grav}}}{2}d(h^2)
+h\mathcal A_xu
&=-g_{\mathrm{grav}}h\,dB-\gamma hu.
\end{aligned}
\end{equation}
Here $d$ acts on the base variable at fixed time, $\nabla$ is the Levi-Civita connection of $g(t)$, and $(\operatorname{div}_g T)_i=(g_x)^{jk}\nabla_jT_{ik}$ for a covariant two-tensor $T$. Every term in the momentum equation is a covector field. 
\end{theorem}
\begin{proof}
Integrating the kinetic equation against $1$ and using \eqref{eq:time-fiber-integral} gives the continuity equation, including $\vartheta h$, by the zeroth and first moment conditions. Integration against $\xi_i$ similarly gives $\partial_t m_i+\vartheta m_i$ and the local momentum balance~\eqref{eq:local-momentum}. The metric-derivative contribution combines with the coordinate divergence to form $\nabla_jK_i{}^j$, as established above. The tensor closure gives $K_i{}^j=hu_i u^j+p(h)\delta_i{}^j$ with $p(h)=g_{\mathrm{grav}}h^2/2$. The force moment is $\int F M\,d\mu_x=-h\mathcal A_xu-\gamma hu$, while the potential contributes $-g_{\mathrm{grav}}h\,dB$ and the collision moments vanish. Finally, $\operatorname{div}_g(pg)=dp=(g_{\mathrm{grav}}/2)d(h^2)$ yields the stated momentum equation.
\end{proof}

This result is conditional on the moment closure; it does not assert that an arbitrary solution of the kinetic equation preserves that closure. 



\subsection{Essential properties of the shallow water system}

For $h>0$, the positive quantity $c=\sqrt{g_{\mathrm{grav}}h}$ is the local gravity-wave speed relative to the fluid, also called the \emph{sound speed} by analogy with barotropic gas dynamics. Here it describes shallow-water surface waves, not acoustic compression of the water.

\begin{proposition}
\label{prop:general-quasilinear}
Let $g(t)$ be the prescribed smooth Riemannian metric and $g_{\mathrm{grav}}>0$. For smooth states in the wet region $h>0$, retain the covector momentum $m=hu$ and write
\[
 U=(h,m_1,\ldots,m_n)^T,\qquad m_i=hu_i,\qquad
 m^j=(g_x)^{jk}m_k,\qquad u^j=(g_x)^{jk}u_k.
\]
Then equation~\eqref{eq:sw-forward} is equivalent in local coordinates to
\begin{equation}\label{eq:general-quasilinear}
 \partial_t U+\sum_{j=1}^n A^j(t,x,U)\partial_jU=S(t,x,U),\qquad
 A^j=\begin{pmatrix}
 0 & (g_x)^{jk}\\
 c^2\delta_i{}^j-u_i u^j & u^j\delta_i{}^k+u_i(g_x)^{jk}
 \end{pmatrix},
\end{equation}
where $\partial_j=\partial/\partial x^j$ and
\begin{equation}\label{eq:shallow-water-sound-speed}
 c^2=p'(h)=g_{\mathrm{grav}}h.
\end{equation}
In the matrix $A^j$, $j$ labels the coordinate direction and is not summed within the displayed matrix; $i,k$ index its momentum rows and columns. The source components are
\begin{equation}\label{eq:general-quasilinear-source}
 \begin{aligned}
 S_h&=(g_x)^{jk}\Gamma^\ell_{jk}m_\ell-\vartheta h,\\
 S_{m_i}&=u_i(g_x)^{jk}\Gamma^\ell_{jk}m_\ell
 +\Gamma^\ell_{ji}u^j m_\ell
 -g_{\mathrm{grav}}h\,\partial_iB-(\mathcal A_xm)_i-\gamma m_i-\vartheta m_i.
 \end{aligned}
\end{equation}
Repeated indices in these source expressions are summed. The time variation of the metric contributes $-\vartheta U$ to the source; it changes the values of the coefficient matrices through $g(t)$ but adds no spatial derivative of $U$. In particular, the spatial variation of the metric is included through the connection; it has not been discarded when extracting the coefficient matrices.

For a nonzero covector $\beta=\beta_jdx^j$, the directional coefficient matrix $A(\beta)=\sum_j\beta_jA^j$ has characteristic polynomial
\[
 \det(\lambda I-A(\beta))
 =(\lambda-u^j\beta_j)^{n-1}
 \left[(\lambda-u^j\beta_j)^2-c^2|\beta|_{g_x^{-1}}^2\right].
\]
Its eigenvalues are
\begin{equation}\label{eq:general-characteristic-speeds}
 u^j\beta_j-c|\beta|_{g_x^{-1}},\qquad
 u^j\beta_j\quad\text{(multiplicity $n-1$)},\qquad
 u^j\beta_j+c|\beta|_{g_x^{-1}}.
\end{equation}
For every nonzero $\beta$, the eigenvectors form a complete basis. Consequently, the system is hyperbolic for $h>0$, with $n-1$ transverse modes (none when $n=1$).
\end{proposition}
\begin{proof}
By metric compatibility, the continuity equation expands as
\[
 \partial_t h+(g_x)^{jk}\partial_jm_k
 =(g_x)^{jk}\Gamma^\ell_{jk}m_\ell-\vartheta h.
\]
For momentum, $\operatorname{div}_g(pg)=dp$ and the product rule give
\[
 (\operatorname{div}_g(hu\otimes u))_i
 =u^j\nabla_jm_i+u_i\nabla_jm^j-u_i u^j\partial_jh.
\]
Indeed, apply the covariant derivative to $m_i m^j/h$ and use $\nabla g=0$. Substituting
\[
 \nabla_jm_i=\partial_jm_i-\Gamma^\ell_{ji}m_\ell,
 \qquad
 \nabla_jm^j=(g_x)^{jk}
 (\partial_jm_k-\Gamma^\ell_{jk}m_\ell)
\]
into the momentum balance and using $\partial_i p=c^2\partial_i h$ yields exactly the matrices~\eqref{eq:general-quasilinear} and sources~\eqref{eq:general-quasilinear-source}. Conversely, regrouping these terms as covariant derivatives recovers~\eqref{eq:sw-forward}.

To verify the characteristic structure, freeze $(t,x)$ and the state and work directly with perturbations $(\delta h,\delta m)$ of the conservative variables. From \eqref{eq:general-quasilinear}, the directional matrix acts as
\[
 A(\beta)\begin{pmatrix}\delta h\\\delta m\end{pmatrix}
 =\begin{pmatrix}
 g_x^{-1}(\beta,\delta m)\\
 \bigl(c^2\beta-(u^j\beta_j)u\bigr)\delta h
 +(u^j\beta_j)\delta m+u\,g_x^{-1}(\beta,\delta m)
 \end{pmatrix}.
\]
If $\delta h=0$ and $g_x^{-1}(\beta,\delta m)=0$, this equals
$(u^j\beta_j)(0,\delta m)^T$. These transverse momentum perturbations form an $(n-1)$-dimensional eigenspace. The two remaining eigenvectors are
\[
 \begin{pmatrix}1\\u\pm c\,\beta/|\beta|_{g_x^{-1}}\end{pmatrix},
\]
where $u=m/h$ is the fixed background covector, not a change of state variables. Direct substitution gives
\[
 A(\beta)\begin{pmatrix}1\\u\pm c\,\beta/|\beta|_{g_x^{-1}}\end{pmatrix}
 =\bigl(u^j\beta_j\pm c|\beta|_{g_x^{-1}}\bigr)
 \begin{pmatrix}1\\u\pm c\,\beta/|\beta|_{g_x^{-1}}\end{pmatrix}.
\]
Since $c>0$ and $\beta\ne0$, these two eigenvalues are distinct from each other and from the transverse eigenvalue. Together with a basis of the transverse eigenspace, their eigenvectors therefore give $n+1$ linearly independent vectors. This proves the characteristic polynomial, the stated multiplicities, and completeness of the eigenvectors, including the case $n=1$, when there are no transverse modes.
\end{proof}

For a unit tangent vector $e$, the characteristic speeds are $u(e)\pm c$ and, when $n>1$, $u(e)$, the latter corresponding to transverse momentum transport.

\begin{corollary}
\label{cor:mass-and-rest}
Under the assumptions of Theorem~\ref{thm:intrinsic-shallow-water}, let $h,u$ be a smooth solution of \eqref{eq:sw-forward}. For a fixed relatively compact region $D\subset\Omega$ with smooth boundary,
\begin{equation}\label{eq:mass-region}
\frac{d}{dt}\int_D h\,dV_g
=-\int_{\partial D}g_x(hu^\sharp,\nu)\,d\sigma_g,
\end{equation}
where $\nu$ and $d\sigma_g$ are the outward unit normal and induced boundary measure for the current metric $g(t)$. If $\Omega$ is compact without boundary, or compact with boundary and zero normal mass flux, then
\begin{equation}\label{eq:total-mass}
\mathcal M(t)=\int_\Omega h\,dV_g
=\int_{T^*\Omega}M\,dL,\qquad
\frac{d\mathcal M}{dt}=0.
\end{equation}
For the rest-state statements only, assume additionally that $g$ is independent of time. Then, for a time-independent solution with $u=0$, each connected wet region satisfies
\begin{equation}\label{eq:lake-at-rest}
\boxed{d(h+B)=0,\qquad h+B=\text{constant}.}
\end{equation}
Conversely, a smooth, time-independent state with $u=0$, $h>0$ and $h+B$ constant solves \eqref{eq:sw-forward} on that region.
\end{corollary}
\begin{proof}
Since $D$ is fixed, differentiation of its time-dependent volume element gives
\[
 \frac{d}{dt}\int_D h\,dV_g=\int_D(\partial_t h+\vartheta h)\,dV_g.
\]
The continuity equation and the divergence theorem for $g(t)$ then give \eqref{eq:mass-region}. Taking $D=\Omega$ gives conservation when the boundary is empty or the normal flux vanishes. The equality of the base and phase-space integrals follows from $\int M\,d\mu_x=h$ and $dL=dV_g\,d\mu_x$.

For a stationary state with $u=0$ and time independent metric, the transport and rotation terms vanish, and the momentum equation reduces to
$
 g_{\mathrm{grav}}h\,d(h+B)=0.
$
On a wet region, division by $g_{\mathrm{grav}}h>0$ gives $d(h+B)=0$, so $h+B$ is constant on each connected component. Conversely, this condition makes the momentum equation vanish, while stationary continuity is automatic for the fixed metric.
\end{proof}
For a time-dependent metric, $u=0$ instead requires $\partial_t h+\vartheta h=0$ together with $d(h+B)=0$ on wet regions. A fixed-depth lake at rest is therefore generally incompatible with a changing volume element.

On a noncompact manifold, global conservation additionally requires finite total mass, justification of differentiation under the integral, and vanishing flux at infinity. The rest-state conclusion above is local to wet regions; it does not require division by $h$ at dry points. In the depth interpretation, $\mathcal M$ is fluid volume; multiplying by a fixed material density gives physical mass.



\begin{proposition}
\label{prop:continuous-energy}\label{prop:viscous-entropy}
Let $g(t)$ be a prescribed smooth Riemannian metric, $B$ smooth and independent
of time, and $\mathcal A_x$ skew-adjoint with respect to $g_x^{-1}$ at each
time, with $\gamma\geq0$ as above. The mechanical energy
\begin{equation}\label{eq:continuous-energy-density}
 \mathcal E(t,x,h,u)=\frac{h}{2}|u|_{g_x^{-1}}^2
       +\frac{g_{\mathrm{grav}}}{2}h^2+g_{\mathrm{grav}}Bh,
 \qquad m=hu,\qquad p=\frac{g_{\mathrm{grav}}}{2}h^2,
\end{equation}
is a strictly convex mathematical entropy in the conservative variables
$(h,m)$ for $h>0$ at each fixed $(t,x)$. Define the geometric work rate by
\begin{equation}\label{eq:metric-work}
 \mathcal W_g=-\frac h2\dot g(u^\sharp,u^\sharp)-p\vartheta.
\end{equation}
The two terms in $\mathcal W_g$ represent work associated with changes in the velocity norm and geometric volume, respectively.
At $h=0$, we take $m=0$ and use the conventions $u=0$ and $\gamma=0$;
the energy, energy flux, geometric work, and frictional dissipation then vanish.
The convergence assumptions below include these dry-state values.

Assume that smooth positive-depth approximations
$(h^\varepsilon,u^\varepsilon)$, with the same prescribed metric $g(t)$,
satisfy the continuity equation and the momentum equation in
\eqref{eq:sw-forward} with the additional right-hand side
$\varepsilon\operatorname{div}_g(h^\varepsilon\nabla u^\varepsilon)$,
where $(\nabla u)_{ij}=\nabla_j u_i$.
Assume that they converge to a weak solution $(h,u)$ of
\eqref{eq:sw-forward} and that their energy densities and fluxes converge to
$\mathcal E$ and $(\mathcal E+p)u^\sharp$ in $L^1_{\mathrm{loc}}$ on space--time.
Also assume
\[
 \gamma^\varepsilon h^\varepsilon|u^\varepsilon|_{g_x^{-1}}^2
 \longrightarrow\gamma h|u|_{g_x^{-1}}^2,
 \qquad \mathcal W_g^\varepsilon\longrightarrow\mathcal W_g
 \quad\text{in }L^1_{\mathrm{loc}},
\]
where $\gamma^\varepsilon=\gamma(x,h^\varepsilon,|u^\varepsilon|_{g_x^{-1}})$
and $\mathcal W_g^\varepsilon$ is \eqref{eq:metric-work} evaluated at the
approximate state, and assume
\[
 \varepsilon h^\varepsilon\operatorname{grad}_g
       \frac{|u^\varepsilon|_{g_x^{-1}}^2}{2}\longrightarrow0
 \quad\text{in distributions},\qquad
 \operatorname{grad}_g a=(da)^\sharp.
\]
Then the limiting weak solution satisfies
\begin{equation}\label{eq:continuous-entropy-inequality}
 \partial_t\mathcal E+\vartheta\mathcal E+
 \operatorname{div}_g\bigl((\mathcal E+p)u^\sharp\bigr)
 \leq\mathcal W_g-\gamma h|u|_{g_x^{-1}}^2
 \quad\text{in distributions}.
\end{equation}
Equality holds for smooth solutions with $h>0$.
\end{proposition}
\begin{proof}
Set $K=\tfrac12h|u|_{g_x^{-1}}^2$. Since
$\partial_t g^{-1}(u,u)=-\dot g(u^\sharp,u^\sharp)$ when $u$ is held fixed,
contracting the smooth momentum equation with $u^\sharp$ and using
continuity, spatial metric compatibility, and
$g_x^{-1}(u,\mathcal A_xu)=0$ gives
\[
 \partial_t K+\vartheta K+\operatorname{div}_g(Ku^\sharp)
 =-dp(u^\sharp)-g_{\mathrm{grav}}h\,dB(u^\sharp)
  -\gamma h|u|_{g_x^{-1}}^2-\frac h2\dot g(u^\sharp,u^\sharp).
\]
Continuity also gives
\[
 \partial_t p+\vartheta p+\operatorname{div}_g(pu^\sharp)
 =-p\operatorname{div}_g u^\sharp-p\vartheta,
\]
and, since $B$ is independent of time,
\[
 \partial_t(g_{\mathrm{grav}}Bh)+\vartheta g_{\mathrm{grav}}Bh
 +\operatorname{div}_g(g_{\mathrm{grav}}Bh u^\sharp)
 =g_{\mathrm{grav}}h\,dB(u^\sharp).
\]
Adding these identities and combining the pressure terms yields
\begin{equation}\label{eq:continuous-energy-law}
 \partial_t\mathcal E+\vartheta\mathcal E+
 \operatorname{div}_g\bigl((\mathcal E+p)u^\sharp\bigr)
 =\mathcal W_g-\gamma h|u|_{g_x^{-1}}^2.
\end{equation}
At fixed $(t,x)$ and for $h>0$, write
$\widetilde{\mathcal E}(t,x,h,m)=\mathcal E(t,x,h,m/h)$. Its second variation is
\[
 D^2_{(h,m)}\widetilde{\mathcal E}[(\delta h,\delta m),(\delta h,\delta m)]
 =\frac1h|\delta m-u\,\delta h|_{g_x^{-1}}^2
       +g_{\mathrm{grav}}(\delta h)^2>0
\]
for every nonzero variation, proving strict convexity.
For the viscous approximations, the same calculation and spatial integration
by parts give
\begin{equation}\label{eq:viscous-energy}
\begin{aligned}
 \partial_t\mathcal E^\varepsilon+\vartheta\mathcal E^\varepsilon+
 \operatorname{div}_g\bigl((\mathcal E^\varepsilon+p^\varepsilon)
                         (u^\varepsilon)^\sharp\bigr)
 &=\mathcal W_g^\varepsilon
 +\varepsilon\operatorname{div}_g\left(
 h^\varepsilon\operatorname{grad}_g
       \frac{|u^\varepsilon|_{g_x^{-1}}^2}{2}\right)\\
 &\quad-\varepsilon h^\varepsilon|\nabla u^\varepsilon|_g^2
 -\gamma^\varepsilon h^\varepsilon|u^\varepsilon|_{g_x^{-1}}^2,
\end{aligned}
\end{equation}
where $|\nabla u|_g^2=(g_x)^{ia}(g_x)^{jb}(\nabla_j u_i)(\nabla_b u_a)$.
Test against any nonnegative
$\varphi\in C_c^\infty((0,T)\times\Omega)$ supported in the spatial interior.
In integration by parts with $dV_{g(t)}\,dt$, the time derivative of the
volume element cancels the term $\vartheta\mathcal E^\varepsilon$.
The viscous flux tends to zero by assumption, and the nonnegative viscous
dissipation can be discarded. Passing to the limit yields
\begin{equation}\label{eq:weak-entropy-test}
 \int_0^T\!\int_\Omega
 \left[\mathcal E\,\partial_t\varphi
       +(\mathcal E+p)\,d\varphi(u^\sharp)
       +(\mathcal W_g-\gamma h|u|_{g_x^{-1}}^2)\varphi\right]dV_g\,dt\geq0,
\end{equation}
which is \eqref{eq:continuous-entropy-inequality}.
\end{proof}
For smooth positive-depth solutions on a compact manifold with no boundary
or an impermeable boundary, integration gives
\[
 \frac{d}{dt}\int_\Omega\mathcal E\,dV_g
 =\int_\Omega\bigl(\mathcal W_g-\gamma h|u|_{g_x^{-1}}^2\bigr)\,dV_g.
\]
The geometric work has no prescribed sign, so total energy need not be
nonincreasing when the metric varies. For a time-independent metric,
$\vartheta=0$ and $\mathcal W_g=0$: total energy is nonincreasing and is
conserved when $\gamma=0$. Here $\mathcal E$ is a mathematical entropy,
not a thermodynamic one.

\section{The sphere}

\begin{figure}[htp]
\centering
\includegraphics[width=.8\textwidth]{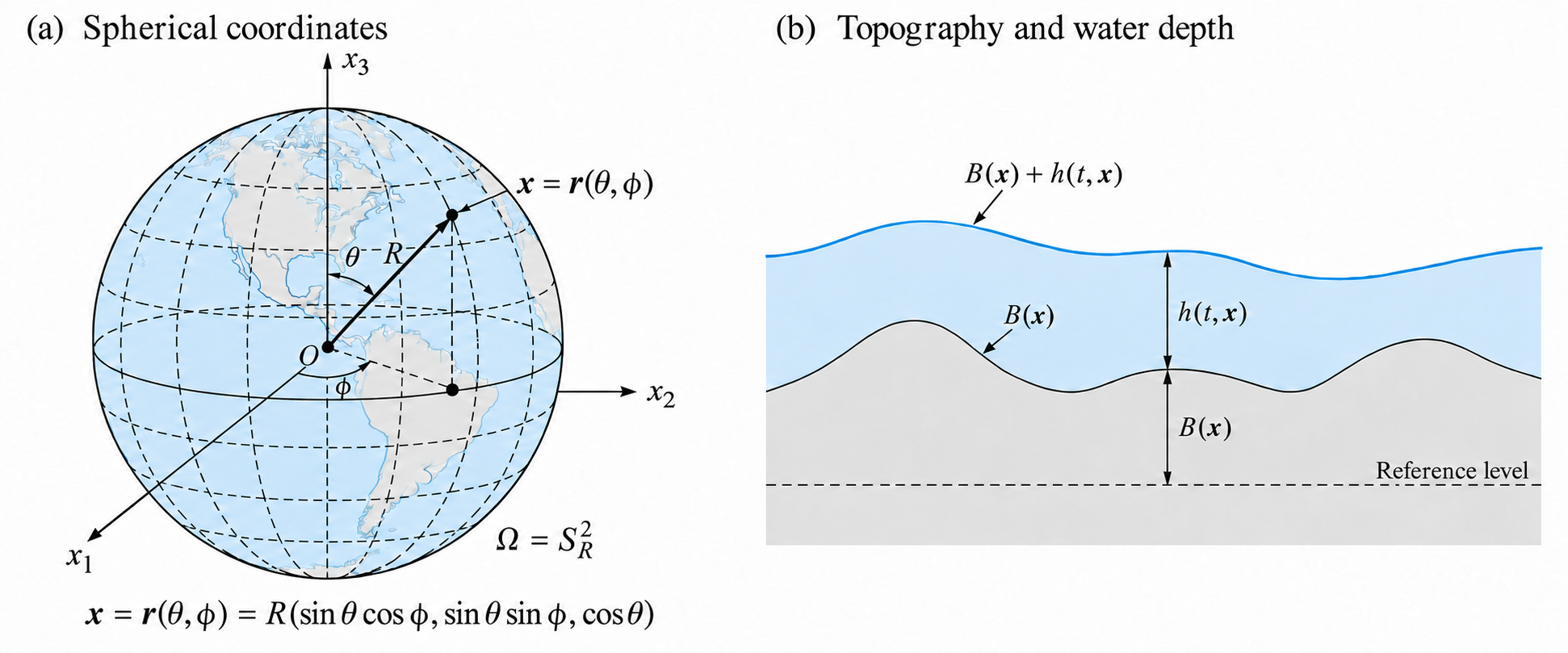}
\caption{Schematic of the ambient space: sphere}
\label{fig:spherical-schematic}
\end{figure}

From this section through the numerical experiments, set $\Omega=S_R^2$ and use the time-independent round metric of constant radius $R>0$ induced from $\mathbb R^3$. Thus $\dot g=0$, $\vartheta=0$, and $\mathcal W_g=0$ throughout the spherical specialization and its discretization. Define
\begin{equation}\label{eq:spherical-parametrization}
x=\mathbf r(\theta,\phi)
=R(\sin\theta\cos\phi,\sin\theta\sin\phi,\cos\theta).
\end{equation}
The colatitude $\theta\in(0,\pi)$ is measured from the positive third axis (the North Pole direction), so it equals $\pi/2$ minus latitude; $\phi$ is azimuth. An open angular branch for $\phi$ excludes a meridian as well as the poles; other charts cover those points. At $x=\mathbf r(\theta,\phi)$,
\[
g_x=R^2(d\theta)_x\otimes(d\theta)_x
+R^2\sin^2\theta\,(d\phi)_x\otimes(d\phi)_x,
\qquad ((g_x)_{ij})=R^2\begin{pmatrix}1&0\\0&\sin^2\theta\end{pmatrix}.
\]
Thus $G(x)=\det((g_x)_{ij})=R^4\sin^2\theta$, $dV_g=R^2\sin\theta\,d\theta\,d\phi$, and for
$\xi=\xi_\theta(d\theta)_x+\xi_\phi(d\phi)_x$,
\begin{equation}\label{eq:spherical-fiber}
|\xi|_{g_x^{-1}}^2=\frac1{R^2}\left(\xi_\theta^2+\frac{\xi_\phi^2}{\sin^2\theta}\right),
\qquad d\mu_x=\frac{d\xi_\theta\,d\xi_\phi}{R^2\sin\theta}.
\end{equation}
Here $\xi_\theta,\xi_\phi$ are the components of $\xi$ in the coordinate
covector basis $(d\theta)_x,(d\phi)_x$; the indices label components, not
derivatives. This basis is orthogonal but not orthonormal: its covectors
have dual-metric lengths $1/R$ and $1/(R\sin\theta)$, respectively.
The same covector written in the orthonormal basis is
\[
\xi=\frac{\xi_\theta}{R}\bigl(R(d\theta)_x\bigr)
   +\frac{\xi_\phi}{R\sin\theta}\bigl(R\sin\theta(d\phi)_x\bigr).
\]
Thus its orthonormal components are $\xi_\theta/R$ and
$\xi_\phi/(R\sin\theta)$. At fixed $x$, the Jacobian of this rescaling
is $1/(R^2\sin\theta)=G(x)^{-1/2}$, so $d\mu_x$ is ordinary Euclidean
area measure in these components.


On this coordinate domain,
\[
\omega=d\theta\wedge d\xi_\theta+d\phi\wedge d\xi_\phi,
\]
\[
\{a,b\}=\frac{\partial a}{\partial\theta}\frac{\partial b}{\partial\xi_\theta}
-\frac{\partial a}{\partial\xi_\theta}\frac{\partial b}{\partial\theta}
+\frac{\partial a}{\partial\phi}\frac{\partial b}{\partial\xi_\phi}
-\frac{\partial a}{\partial\xi_\phi}\frac{\partial b}{\partial\phi},
\qquad
H=\frac1{2R^2}\left(\xi_\theta^2+\frac{\xi_\phi^2}{\sin^2\theta}\right)
+g_{\mathrm{grav}}B.
\]
Use the outward orientation and let $\boldsymbol\varpi$ be planetary angular velocity, avoiding a conflict with the manifold symbol $\Omega$. With $n(x)=x/R$, rotation acts on the tangent vector $\xi^\sharp$ by the cross product with $n(x)$; lowering the result gives its action on the covector $\xi$:
\[
\mathcal J_x\xi=\bigl(n(x)\times\xi^\sharp\bigr)^\flat
=-\frac{\xi_\phi}{\sin\theta}(d\theta)_x
+\sin\theta\,\xi_\theta(d\phi)_x,
\qquad \mathcal A_x=f(x)\mathcal J_x.
\]
Here $f(x)=2\boldsymbol\varpi\cdot n(x)$; for $\boldsymbol\varpi=\varpi(0,0,1)$, $f=2\varpi\cos\theta$. The force $-\mathcal A_x\xi$ is the covector form of the tangential Coriolis acceleration. Writing $\widehat M$ for the coordinate representation, the kinetic equation is
\begin{align*}
\frac{\partial\widehat M}{\partial t}
+\frac{\xi_\theta}{R^2}\frac{\partial\widehat M}{\partial\theta}
+\frac{\xi_\phi}{R^2\sin^2\theta}\frac{\partial\widehat M}{\partial\phi}
&+\left(\frac{\xi_\phi^2\cos\theta}{R^2\sin^3\theta}
-g_{\mathrm{grav}}\frac{\partial B}{\partial\theta}
+\frac{f\xi_\phi}{\sin\theta}-\gamma u_\theta\right)
\frac{\partial\widehat M}{\partial\xi_\theta}\\
&+\left(-g_{\mathrm{grav}}\frac{\partial B}{\partial\phi}
-f\sin\theta\,\xi_\theta-\gamma u_\phi\right)
\frac{\partial\widehat M}{\partial\xi_\phi}=\widehat C.
\end{align*}
The poles are coordinate singularities, not singularities of the Hamiltonian, the force, or the bracket.


In the orthonormal components, the conditions \eqref{eq:equilibrium-targets} become
\begin{align*}
\int M\,d\mu_x&=h,
\int\frac{\xi_\theta}{R}M\,d\mu_x =h\frac{u_\theta}{R},\qquad
\int\frac{\xi_\phi}{R\sin\theta}M\,d\mu_x=h\frac{u_\phi}{R\sin\theta},\\
\int\left(\frac{\xi_\theta}{R}\right)^2M\,d\mu_x
&=h\left(\frac{u_\theta}{R}\right)^2+p,
\int\frac{\xi_\theta}{R}\frac{\xi_\phi}{R\sin\theta}M\,d\mu_x
=h\frac{u_\theta}{R}\frac{u_\phi}{R\sin\theta},\\
\int\left(\frac{\xi_\phi}{R\sin\theta}\right)^2M\,d\mu_x
&=h\left(\frac{u_\phi}{R\sin\theta}\right)^2+p,
\qquad p=\tfrac12g_{\mathrm{grav}}h^2.
\end{align*}
Here all integrals are over $T_x^*S_R^2$; in coordinates,
$M\,d\mu_x=\widehat M\,d\xi_\theta d\xi_\phi/(R^2\sin\theta)$.
Dividing the two diagonal moment identities by $R^2$ and
$R^2\sin^2\theta$, respectively, and adding gives
\[
\int\left[\left(\frac{\xi_\theta}{R}\right)^2
+\left(\frac{\xi_\phi}{R\sin\theta}\right)^2\right]M\,d\mu_x
=h\left[\left(\frac{u_\theta}{R}\right)^2
+\left(\frac{u_\phi}{R\sin\theta}\right)^2\right]+g_{\mathrm{grav}}h^2.
\]
These identities are satisfied, in particular, by the intrinsic equilibrium \eqref{eq:general-gibbs-equilibrium}. Since the sphere is two-dimensional, $n=2$, and hence
\begin{equation}\label{eq:spherical-gibbs-equilibrium}
M(t,x,\xi)=\frac{3}{5\pi g_{\mathrm{grav}}}
\sqrt{\max\left(1-
\frac{2|\xi-u(t,x)|_{g_x^{-1}}^2}{5g_{\mathrm{grav}}h(t,x)},0\right)}.
\end{equation}
Equivalently, in the spherical coordinates used above,
\begin{equation}\label{eq:spherical-gibbs-coordinate}
\widehat M(t,\theta,\phi,\xi_\theta,\xi_\phi)
=\frac{3}{5\pi g_{\mathrm{grav}}}
\sqrt{\max\left(1-
\frac{2}{5g_{\mathrm{grav}}h}
\left[
\frac{(\xi_\theta-u_\theta)^2}{R^2}
+\frac{(\xi_\phi-u_\phi)^2}{R^2\sin^2\theta}
\right],0\right)}.
\end{equation}
Its support is the metric ball
$
|\xi-u|_{g_x^{-1}}\leq \sqrt{\frac52g_{\mathrm{grav}}h}.
$
The apparent anisotropy of the support ellipse in the coordinate components $(\xi_\theta,\xi_\phi)$ is a coordinate effect: in the orthonormal components $(\xi_\theta/R,\xi_\phi/(R\sin\theta))$ it is an ordinary Euclidean disk. 

\subsection{Spherical fluid equations and geometric terms}
The nonzero Christoffel symbols are
$
\Gamma^\theta_{\phi\phi}=-\sin\theta\cos\theta,\qquad
\Gamma^\phi_{\theta\phi}=\Gamma^\phi_{\phi\theta}=\cot\theta.
$
Substituting these Christoffel symbols into the general quasilinear source
\eqref{eq:general-quasilinear-source} yields the geometric terms containing
$\cot\theta$ in \eqref{eq:sphere-quasilinear-source} below.

\paragraph{Quasilinear system in covector momentum.}
To specialize \eqref{eq:general-quasilinear}, use the coordinate order $(\theta,\phi)$ and the state
\[
 U=(h,m_\theta,m_\phi)^T,\qquad
 m_\theta=hu_\theta,\quad m_\phi=hu_\phi,\qquad
 u^\theta=\frac{u_\theta}{R^2},\quad
 u^\phi=\frac{u_\phi}{R^2\sin^2\theta}.
\]
Here $u^\theta,u^\phi$ are called contravariant components because they are
the coefficients of the tangent vector
$u^\sharp=u^\theta\partial_\theta+u^\phi\partial_\phi$ in the coordinate basis.
They are obtained from the covector components $u_\theta,u_\phi$ by raising
the index with the inverse metric. They represent angular transport rates;
the physical velocity components are $u_\theta/R,u_\phi/(R\sin\theta)$.
On $0<\theta<\pi$ and for $h>0$,
\begin{equation}\label{eq:sphere-quasilinear}
 \partial_t U+A^\theta\partial_\theta U+A^\phi\partial_\phi U=S,
\end{equation}
with $c^2=g_{\mathrm{grav}}h$ and
\begin{equation}\label{eq:sphere-quasilinear-matrices}
 \begin{aligned}
 A^\theta&=\begin{pmatrix}
 0& R^{-2}&0\\
 c^2-u_\theta u^\theta&2u^\theta&0\\
 -u_\phi u^\theta&u_\phi/R^2&u^\theta
 \end{pmatrix}, 
 A^\phi&=\begin{pmatrix}
 0&0&(R^2\sin^2\theta)^{-1}\\
 -u_\theta u^\phi&u^\phi&u_\theta/(R^2\sin^2\theta)\\
 c^2-u_\phi u^\phi&0&2u^\phi
 \end{pmatrix}.
 \end{aligned}
\end{equation}
The source in this coordinate basis is
\begin{equation}\label{eq:sphere-quasilinear-source}
 S=\begin{pmatrix}
 -m_\theta\cot\theta/R^2\\[1mm]
 \displaystyle\frac{\cot\theta}{R^2}
 \left(-u_\theta m_\theta+\frac{u_\phi m_\phi}{\sin^2\theta}\right)
 -g_{\mathrm{grav}}h\,\partial_\theta B+\frac{f m_\phi}{\sin\theta}-\gamma m_\theta\\[1mm]
 -u_\theta m_\phi\cot\theta/R^2
 -g_{\mathrm{grav}}h\,\partial_\phi B-f\sin\theta\,m_\theta-\gamma m_\phi
 \end{pmatrix}.
\end{equation}
The Coriolis signs follow from $\mathcal A_x=f\mathcal J_x$ above. The pressure terms $c^2\partial_i h$ contribute the $c^2$ entries in the coefficient matrices $A^\theta$ and $A^\phi$; metric compatibility eliminates any additional pressure source in this covector formulation. The geometric source therefore differs in appearance from the source obtained when evolving physical momentum components.

The eigenvalues of the two coordinate matrices are respectively
\[
 \operatorname{spec}(A^\theta)=
 \left\{u^\theta-\frac cR,\ u^\theta,\ u^\theta+\frac cR\right\},\qquad
 \operatorname{spec}(A^\phi)=
 \left\{u^\phi-\frac c{R\sin\theta},\ u^\phi,\ u^\phi+\frac c{R\sin\theta}\right\}.
\]
These are angular rates. Multiplying by the respective metric scale factors $R$ and $R\sin\theta$ recovers the physical normal velocities plus $-c,0,c$. The apparent singularities at the poles belong to the angular chart; the intrinsic formulation remains valid in regular charts. The physical-component equations below are equivalent to \eqref{eq:sphere-quasilinear} after the spatially varying change of momentum components.


To express the fluid equations using physical velocity components, set
\[
e_\theta=\frac1R\frac{\partial\mathbf r}{\partial\theta},\qquad
e_\phi=\frac1{R\sin\theta}\frac{\partial\mathbf r}{\partial\phi},\qquad
u^\sharp=\frac{u_\theta}{R}e_\theta+\frac{u_\phi}{R\sin\theta}e_\phi.
\]
Here $u_\theta/R$ is southward and $u_\phi/(R\sin\theta)$ eastward. Then \eqref{eq:sw-forward} is equivalent, for $h>0$, to
\begin{align*}
&\frac{\partial h}{\partial t}
+\frac1{R^2\sin\theta}\frac{\partial}{\partial\theta}(\sin\theta\,h u_\theta)
+\frac1{R^2\sin\theta}\frac{\partial}{\partial\phi}
 \left(h\frac{u_\phi}{\sin\theta}\right)=0,\\[4mm]
&\frac{\partial(h u_\theta/R)}{\partial t}
+\frac1{R\sin\theta}\frac{\partial}{\partial\theta}
 \left(\sin\theta\,h\left(\frac{u_\theta}{R}\right)^2\right)
+\frac1{R\sin\theta}\frac{\partial}{\partial\phi}
 \left(h\frac{u_\theta}{R}\frac{u_\phi}{R\sin\theta}\right)
+\frac1R\frac{\partial p}{\partial\theta}\\[1mm]
&\qquad=\frac hR\left(\frac{u_\phi}{R\sin\theta}\right)^2\cot\theta
+fh\frac{u_\phi}{R\sin\theta}
-\frac{g_{\mathrm{grav}}h}{R}\frac{\partial B}{\partial\theta}-\gamma h\frac{u_\theta}{R},\\[4mm]
&\frac{\partial\left(h\frac{u_\phi}{R\sin\theta}\right)}{\partial t}
+\frac1{R\sin\theta}\frac{\partial}{\partial\theta}
 \left(\sin\theta\,h\frac{u_\theta}{R}\frac{u_\phi}{R\sin\theta}\right)
+\frac1{R\sin\theta}\frac{\partial}{\partial\phi}
 \left(h\left(\frac{u_\phi}{R\sin\theta}\right)^2\right)
+\frac1{R\sin\theta}\frac{\partial p}{\partial\phi}\\[1mm]
&\qquad=-\frac hR\frac{u_\theta}{R}\frac{u_\phi}{R\sin\theta}\cot\theta
-fh\frac{u_\theta}{R}
-\frac{g_{\mathrm{grav}}h}{R\sin\theta}\frac{\partial B}{\partial\phi}-\gamma h\frac{u_\phi}{R\sin\theta}.
\end{align*}
The terms involving $\cot\theta$ arise by differentiating the moving basis; the factor $\sin\theta$ in continuity comes from area. Replacing planar derivatives by angular derivatives while treating the basis as constant omits these terms. A correct coordinate calculation agrees with the intrinsic equations. A coordinate change of a flat metric cannot create the sphere's nonzero curvature.

\section{A two-dimensional spherical kinetic finite-volume scheme}
\label{sec:ps-construction}

We now derive a kinetic scheme for the shallow water equations on the sphere with its fixed round metric, based on the work of Perthame and Simeoni~\cite{perthame2001}, hereafter PS. References to PS equations below refer to their numbering, not ours. This section discretizes the classical shallow-water equations of Section~4, independently of the semiclassical construction in Appendix~\ref{app:semiclassical}. The formulas below include bottom friction, with the Manning implementation specified in Section~\ref{sec:numerical-results}. The potential-jump reflection rule in PS is replaced by the hydrostatic interface treatment, as it will be explained below. 



Similar to the intervals $C_i$ in PS, here we consider an angular cell on $S_R^2$,
\[
 K_{ij}=\{(\phi,\theta):\phi_{i-1/2}<\phi<\phi_{i+1/2},\;
             \theta_{j-1/2}<\theta<\theta_{j+1/2}\}.
\]
Here, $i$ indexes longitude and $j$ indexes colatitude. The uniform angular widths are $\Delta\phi=\phi_{i+1/2}-\phi_{i-1/2}$ and $\Delta\theta=\theta_{j+1/2}-\theta_{j-1/2}$; $(\phi_i,\theta_j)$ are the angular midpoints and $\theta_K=\theta_j$ for $K=K_{ij}$. Denote its east, west, south and north faces by $E,W,S,N$, respectively. The physical area and edge measures are
\begin{equation}\label{eq:ps-measures}
 \begin{aligned}
 A_j&=R^2\Delta\phi\bigl(\cos\theta_{j-1/2}-\cos\theta_{j+1/2}\bigr),
 L_\phi&=R\Delta\theta,\qquad
 L_{\theta,j\pm1/2}=R\Delta\phi\sin\theta_{j\pm1/2}.
 \end{aligned}
\end{equation}
For any face $f\subset\partial K$, define $L_f$ to be its physical arclength, not its angular width. More explicitly, for $K=K_{ij}$,
\begin{equation}\label{eq:ps-face-length}
 L_f=\int_f\mathrm ds_g=
 \begin{cases}
 L_\phi=R\Delta\theta,&f=E\text{ or }W,\\
 L_{\theta,j+1/2}=R\sin\theta_{j+1/2}\Delta\phi,&f=S,\\
 L_{\theta,j-1/2}=R\sin\theta_{j-1/2}\Delta\phi,&f=N.
 \end{cases}
\end{equation}
Here $\mathrm ds_g$ is the line element induced by the spherical metric. The notation $A_K=A_j$ means the physical area of $K$, and $\sum_{f\subset\partial K}$ sums over its four faces. A shared face has the same $L_f$ for both neighboring cells; polar faces have length zero. The subscript $\phi$ labels a face of constant longitude, whose tangent runs along a meridian; the subscript $\theta$ labels a face of constant colatitude, whose tangent runs along a parallel.
Thus PS's ratio $\Delta t/\Delta x$ is replaced at each edge by $\Delta t L_f/A_j$.  Let us write $U_K^k=(h_K^k,q_{E,K}^k,q_{S,K}^k)^T$ for the stored cell means at time $t^k$, where $k$ is the time-level index, with $q_E=h\velphi$, $q_S=h\veltheta$, and let $B_K$ be the discrete bottom. Conceptually these are means with respect to $R^2\sin\theta\dd\theta\dd\phi$. The Earth setup initializes them by sampling the prescribed fields at cell centers. The code also samples its interpolated topography there and uses that same discrete bottom throughout. 


For reference the physical fluxes, expressed in the stored component order, are
\begin{equation}\label{eq:ps-physical-fluxes}
 F_\phi(U)=\begin{pmatrix}h\velphi\\h\left(\velphi\right)^2+p\\h\velphi\veltheta\end{pmatrix},\qquad
 F_\theta(U)=\begin{pmatrix}h\veltheta\\h\velphi\veltheta\\h\left(\veltheta\right)^2+p\end{pmatrix},
 \qquad p=\frac{\ggra h^2}{2}.
\end{equation}
Write $B_\phi=\partial_\phi B$ and $B_\theta=\partial_\theta B$ for bottom derivatives, and $f(\theta)=2\varpi\cos\theta$ for the Coriolis parameter, where $\varpi$ is the rotation rate. This scalar $f(\theta)$ is distinguished from the face index $f$ in $L_f$ and face sums. The equations being discretized are
\begin{equation}\label{eq:ps-physical-balance}
 U_t+\frac{\partial_\phi F_\phi+\partial_\theta(\sin\theta F_\theta)}
                   {R\sin\theta}
 =\begin{pmatrix}
 0\\
 -\ggra hB_\phi/(R\sin\theta)-h\velphi\veltheta\cot\theta/R-fh\veltheta\\
 -\ggra hB_\theta/R+(h\left(\velphi\right)^2+p)\cot\theta/R+fh\velphi
 \end{pmatrix}-\gamma\begin{pmatrix}0\\q_E\\q_S\end{pmatrix}.
\end{equation}



Here $\theta$ is colatitude, $\phi$ is longitude, and $e_\theta,e_\phi$ are the southward and eastward unit tangent vectors. The velocity $u=u_\theta\,\mathrm d\theta+u_\phi\,\mathrm d\phi$ is a covector, as in the preceding sections; its physical components are $u(e_\theta)=u_\theta/R$ and $u(e_\phi)=u_\phi/(R\sin\theta)$. For the particle covector $\xi$, the physical components are $\xi(e_\theta)=\xi_\theta/R$ and $\xi(e_\phi)=\xi_\phi/(R\sin\theta)$.
At faces of constant longitude, choose the normal and transverse directions as $(e_\perp,e_\parallel)=(e_\phi,e_\theta)$. At faces of constant colatitude, choose them as $(e_\perp,e_\parallel)=(e_\theta,e_\phi)$. These normals point in the direction of increasing longitude and colatitude, respectively. They specify a common orientation of each shared face, not different kinds of cells. We write $\xi_\perp=\xi(e_\perp)$ and $\xi_\parallel=\xi(e_\parallel)$ only when expressing face fluxes; in either orthonormal frame $\dd\mu_x=\dd\xi_\perp\dd\xi_\parallel$. Throughout this section $f_j=2\varpi\cos\theta_j$. In a dry cell $h=0$ the momentum and reconstructed physical velocities are set to zero; for $h>0$ the latter are recovered as $q_E/h,q_S/h$. This division is not regularized. Instead, all reconstruction slopes are set to zero on stencils containing a depth at most $h_{\rm rec}$ (default $10^{-2}\,\mathrm m$). 



Using the particle covector $\xi$ and the fiber measure $\dd\mu_x$ of the preceding sections, the local equilibrium at the center $x$ of cell $K$ is
\begin{equation}\label{eq:ps-local-M}
 M_K^k(\xi)=\frac{3}{5\pi\ggra}
 \sqrt{\max\left(1-\frac{2|\xi-u_K^k|_{g_x^{-1}}^2}{5\ggra h_K^k},0\right)}.
\end{equation}
Here $k$ denotes the time level, leaving $m=hu$ for discharge and $n$ for the manifold dimension. Set $M_K^k=0$ for $h_K^k=0$. Its moments, in the stored east/south component order, satisfy
\begin{equation}\label{eq:ps-moment-map}
 \begin{aligned}
 \int_{T_x^*S_R^2}\begin{pmatrix}1\\\xi(e_\phi)\\\xi(e_\theta)\end{pmatrix}
 M_K^k(\xi)\dd\mu_x&=U_K^k,\\
 \int_{T_x^*S_R^2}\xi(e_r)\begin{pmatrix}1\\\xi(e_\phi)\\\xi(e_\theta)\end{pmatrix}
 M_K^k(\xi)\dd\mu_x&=F_r(U_K^k),\qquad r\in\{\phi,\theta\}.
 \end{aligned}
\end{equation}
These are the spherical counterparts of PS's particle-to-fluid moment map. As in the explanation following PS~(17), no collision equation is integrated: after updating the fluid moments, the next stage constructs equilibria from those moments. Here even the distributions need not be stored; their half-space moments are evaluated analytically.


Using the face orientations defined above, label the adjacent states $L,R$ as west/east for a longitude face and north/south for a colatitude face, irrespective of which cell is being updated. In the following formula, $M(h,u(e_\perp),u(e_\parallel);\xi_\perp,\xi_\parallel)$ denotes the equilibrium~\eqref{eq:ps-local-M} expressed in the common face frame with the indicated depth and fluid velocity; $\mathbf1_A$ is one when condition $A$ holds and zero otherwise. The upwind population for homogeneous transport is
\begin{equation}\label{eq:ps-upwind-population}
 M_f^{\rm up}(\xi_\perp,\xi_\parallel)
 =\mathbf1_{\xi_\perp>0}M(h_L,u_L(e_\perp),u_L(e_\parallel);\xi_\perp,\xi_\parallel)
  +\mathbf1_{\xi_\perp<0}M(h_R,u_R(e_\perp),u_R(e_\parallel);\xi_\perp,\xi_\parallel).
\end{equation}
Only the sign of the normal velocity selects the donor. In local
$(h,h u(e_\perp),h u(e_\parallel))$ order its flux is
\begin{equation}\label{eq:ps-integral-flux}
 \mathcal F_f=\int_{\mathbb R^2}\xi_\perp
       \begin{pmatrix}1\\\xi_\perp\\\xi_\parallel\end{pmatrix}
       M_f^{\rm up}(\xi_\perp,\xi_\parallel)\dd \xi_\perp\dd\xi_\parallel.
\end{equation}
Integrating out $\xi_\parallel$ yields the normal marginal, for $h>0$,
\begin{equation}\label{eq:ps-marginal}
 M_\perp(h,u(e_\perp);\xi_\perp)=\frac{3h}{4a}
 \max\left(1-\frac{(\xi_\perp-u(e_\perp))^2}{a^2},0\right),
 \qquad a=\sqrt{5\ggra h/2}.
\end{equation}
The square root disappears because this is an integral over the transverse velocity, not a restriction to zero transverse velocity. Indeed, for $z=\xi_\perp-u(e_\perp)$ with $|z|<a$, setting $w=\xi_\parallel-u(e_\parallel)$ and $b^2=a^2-z^2$ gives
\[
M_\perp=\frac{3}{5\pi\ggra a}\int_{-b}^{b}\sqrt{b^2-w^2}\,dw
=\frac{3b^2}{10\ggra a},
\]
since the integral is the area of a semicircle. Thus the marginal has a parabolic profile.
Its zeroth, first and second full moments are $h$, $hu(e_\perp)$ and $hu(e_\perp)^2+p(h)$; the transverse first moment of $M$ at fixed $\xi_\perp$ is $u(e_\parallel)M_\perp$. Define the positive half moments
\begin{equation}\label{eq:ps-half-moments}
 T(h,u(e_\perp))=\int_0^\infty\xi_\perp M_\perp\,\mathrm d\xi_\perp,
 \qquad P(h,u(e_\perp))=\int_0^\infty\xi_\perp^2M_\perp\,\mathrm d\xi_\perp.
\end{equation}
Both vanish for $h=0$, or for $u(e_\perp)\leq-a$. For $u(e_\perp)\geq a$ they equal $hu(e_\perp)$ and $hu(e_\perp)^2+p(h)$, respectively. When $h>0$ and $-a<u(e_\perp)<a$, write $\chi=1+u(e_\perp)/a\in(0,2)$. The exact expressions are
\[
 T=\frac{3ha}{4}\chi^3\left(\frac{2-\chi}{6}+\frac{\chi}{12}\right),\qquad
 P=\frac{3ha^2}{4}\chi^4\left(\frac{2-\chi}{12}+\frac{\chi}{20}\right).
\]
The local flux is
\begin{equation}\label{eq:ps-kinflux}
 \mathcal F_f=\begin{pmatrix}
 T(h_L,u_L(e_\perp))-T(h_R,-u_R(e_\perp))\\ P(h_L,u_L(e_\perp))+P(h_R,-u_R(e_\perp))\\
 u_L(e_\parallel)T(h_L,u_L(e_\perp))-u_R(e_\parallel)T(h_R,-u_R(e_\perp))
 \end{pmatrix}.
\end{equation} 
Unlike a one-dimensional fluid, the spherical fluid also transports transverse momentum.

To make the connection with PS~(17) explicit, consider the homogeneous, piecewise-constant transport building block in a frozen physical frame. Its formal particle update would be
\begin{equation}\label{eq:ps-transport-step}
 M_K^{\rm tr}(\xi)=M_K^k(\xi)
 -\frac{\Delta t}{A_K}\sum_{f\subset\partial K}
       \sigma_{Kf}L_f \xi_{\perp,f} M_f^{\rm up}(\xi),
\end{equation}
where the superscript $\mathrm{tr}$ denotes the formal transport stage,
$\sigma_{Kf}=1$ on east/south edges and $-1$ on west/north edges, and $\xi_{\perp,f}=\xi(e_{\perp,f})$ is coordinate-oriented, not outward-oriented. In this formal frozen-frame expression, $M_f^{\rm up}(\xi)$ means the face population evaluated at $(\xi(e_{\perp,f}),\xi(e_{\parallel,f}))$. Taking moments produces the four-edge flux difference. Formula~\eqref{eq:ps-transport-step} explains that building block only: the present method updates moments directly, with the bed and geometric corrections below. 


PS~(21)--(22) treat bottom steps through particle reflection and transmission. The present scheme does not use these mechanisms. Instead, it uses hydrostatic reconstruction~\eqref{eq:ps-interface-depths} and pressure corrections~\eqref{eq:ps-sided-flux}, balanced with the discrete source terms to preserve lakes at rest. Our equilibrium differs from that of PS and, at rest, is not solely a function of particle energy; their microscopic equilibrium argument therefore does not apply directly.



Let $Q_K=(\surface_K,h_K,u_{\phi,K}/(R\sin\theta_K),u_{\theta,K}/R)$, with $\surface_K=h_K+B_K$. For either coordinate direction $r$, denote the adjacent cells by $K-r,K+r$ and define componentwise limited increments
\begin{equation}\label{eq:ps-slope}
 D_rQ_K=\operatorname{minmod}\left(
 \beta(Q_K-Q_{K-r}),\frac{Q_{K+r}-Q_{K-r}}2,
 \beta(Q_{K+r}-Q_K)\right),\qquad \beta=1.5.
\end{equation}
The minmod function returns the common sign times the smallest absolute argument when all arguments have the same sign, and zero otherwise. Its values here are cell differences, not derivatives with respect to angle or physical arclength. Replace $D_rh_K$ by $\operatorname{sgn}(D_rh_K)\min(|D_rh_K|,2h_K)$ to ensure nonnegative reconstructed depths at both cell faces. Set $D_r\surface_K=0$ whenever any of the three stencil depths is zero. Finally, set all four components of $D_rQ_K$ to zero if any stencil depth is at most $h_{\rm rec}$; the default is $h_{\rm rec}=10^{-2}\,\mathrm m$. This local first-order reconstruction modifies no cell mean. The reconstructed face values are
\[
 Q_{KE,KW}=Q_K\pm\tfrac12D_\phi Q_K,\qquad
 Q_{KS,KN}=Q_K\pm\tfrac12D_\theta Q_K,
 \qquad B_{Kf}=\surface_{Kf}-h_{Kf}.
\]
The angular mesh is uniform, covers the full sphere, and has an even number of longitude cells. Longitude ghost cells are periodic. Across a pole, ghost data are taken from the opposite meridian $\phi+\pi$: scalars keep their sign and both physical momentum components change sign. Polar face fluxes are set to zero, consistently with their vanishing lengths. These boundary data are refreshed between stages.

From these reconstructed face values form the hydrostatic cut
\begin{equation}\label{eq:ps-interface-depths}
 B_* =\max(B_L,B_R),\qquad
 h_L^*=\max(\surface_L-B_*,0),\qquad h_R^*=\max(\surface_R-B_*,0).
\end{equation}
Use these cut depths in the two populations in
\eqref{eq:ps-upwind-population}, retaining their reconstructed velocities.
Denote the resulting flux, in stored component order, by $F_f^*$.
For each adjacent cell $K$, let $h_{Kf}$ be its reconstructed face depth before the hydrostatic cut and
$h_{Kf}^*$ its face depth after the hydrostatic cut. With $\mathbf b_\phi=(0,1,0)^T$ and $\mathbf b_\theta=(0,0,1)^T$ denoting the corresponding momentum slots (not geometric tangent vectors),
define the cell-side corrected flux, with $r=\phi$ for $E,W$ faces and $r=\theta$ for $S,N$ faces,
\begin{equation}\label{eq:ps-sided-flux}
 \widehat F_{Kf}=F_f^*+
       \bigl[p(h_{Kf})-p(h_{Kf}^*)\bigr]\mathbf b_r.
\end{equation}
There are two cell-side momentum fluxes at one physical edge, just as in
PS~(19)--(20), but their correction is hydrostatic here. Their mass
components are identical.


Let $D_\phi,D_\theta$ be the limited  increments defined in
\eqref{eq:ps-slope}, and set $D_rB=D_r\surface-D_rh$. Define
\[
 \kappa_j=\frac{L_{\theta,j+1/2}-L_{\theta,j-1/2}}{A_j},\qquad
 \overline L_{\theta,j}=\frac{L_{\theta,j+1/2}+L_{\theta,j-1/2}}2,
 \qquad
 \overline p_{K,\theta}=\frac{p(h_{KS})+p(h_{KN})}{2}.
\]
For an input state $U$ and a stage step $\tau>0$, one Euler update is
\begin{equation}\label{eq:ps-euler}
 U_K^{\rm E}=U_K-\frac{\tau}{A_j}
 \left\{L_\phi(\widehat F_{KE}-\widehat F_{KW})
       +L_{\theta,j+1/2}\widehat F_{KS}
       -L_{\theta,j-1/2}\widehat F_{KN}\right\}
       +\tau\,S_K^{\rm raw},
\end{equation}
where $h_K^{\rm E}$ denotes the depth component of $U_K^{\rm E}$; the
superscript $\rm E$ stands for Euler. All states and velocities on the
right-hand side are evaluated at the beginning of that stage, and
\begin{equation}\label{eq:ps-raw-source}
 S_K^{\rm raw}=\begin{pmatrix}
 0\\
 -\ggra h_KD_\phi B_K L_\phi/A_j-h_K\frac{u_{\phi,K}}{R\sin\theta_K}\frac{u_{\theta,K}}{R}\kappa_j-f_jh_K\frac{u_{\theta,K}}{R}\\
 -\ggra h_KD_\theta B_K\overline L_{\theta,j}/A_j
 +(h_K\left(\frac{u_{\phi,K}}{R\sin\theta_K}\right)^2+\overline p_{K,\theta})\kappa_j+f_jh_K\frac{u_{\phi,K}}{R\sin\theta_K}
 \end{pmatrix}-\gamma_K\begin{pmatrix}0\\q_{E,K}\\q_{S,K}\end{pmatrix}.
\end{equation}
The first term in each momentum source is the internal reconstructed bed
force. The other terms come from curvature, rotation, and bottom friction.
Here $\gamma_K=\gamma(x_K,h_K,\sqrt{(q_{E,K}/h_K)^2+(q_{S,K}/h_K)^2})$
for $h_K>0$, and $\gamma_K=0$ in dry cells. The rate is
recomputed from the cell means at each Euler stage. Thus, unlike PS's
one-dimensional construction, the present method has both interface bed
corrections and explicitly assembled cell sources.

The following rearrangement of \eqref{eq:ps-euler} makes the discrete balance
between pressure, topography, and spherical geometry explicit. Write
$\widehat F_{Kf}=\mathcal R_{Kf}+p(h_{Kf})\mathbf b_r$, with
$\mathcal R_{Kf}=F_f^*-p(h_{Kf}^*)\mathbf b_r$.
Symmetric reconstructed face depths $h_K\pm D_rh_K/2$ give
\begin{align}\label{eq:ps-pressure-cancellation}
 p(h_{KE})-p(h_{KW})&=\ggra h_KD_\phi h_K,\\
 \frac{L_{\theta,j+1/2}p(h_{KS})-L_{\theta,j-1/2}p(h_{KN})}{A_j}
 &=\frac{\ggra h_KD_\theta h_K\overline L_{\theta,j}}{A_j}
       +\kappa_j\overline p_{K,\theta}.
\end{align}
Substitution into \eqref{eq:ps-euler} cancels the metric pressure in
\eqref{eq:ps-raw-source} and combines $D_rh+D_rB=D_r\surface$.
To express the time-stepping scheme compactly, define the spatial update
operator $\mathcal L$ and the forward Euler step $E_\tau$ by the resulting
residual formulation
\begin{equation}\label{eq:ps-residual-operator}
 \mathcal L(U)_K=-\frac1{A_K}\sum_{f\subset\partial K}
 \sigma_{Kf}L_f\mathcal R_{Kf}+S_K^{\rm res},
 \qquad E_{\tau}(U)=U+\tau\,\mathcal L(U),
\end{equation}
where $\mathcal L$ is independent of the Euler step size and
\begin{equation}\label{eq:ps-residual-source}
 S_K^{\rm res}=\begin{pmatrix}
 0\\
 -\ggra h_KD_\phi\surface_K L_\phi/A_j
 -h_K\frac{u_{\phi,K}}{R\sin\theta_K}\frac{u_{\theta,K}}R\kappa_j-f_jq_{S,K}\\
 -\ggra h_KD_\theta\surface_K\overline L_{\theta,j}/A_j
 +h_K\left(\frac{u_{\phi,K}}{R\sin\theta_K}\right)^2\kappa_j+f_jq_{E,K}
 \end{pmatrix}-\gamma_K\begin{pmatrix}0\\q_{E,K}\\q_{S,K}\end{pmatrix}.
\end{equation}
In this formulation, the combined pressure and topographic terms in
$S_K^{\rm res}$ depend on discrete differences of the free-surface elevation
$\surface=h+B$. At a lake at rest these differences vanish, as do the
velocity terms and interface residuals, yielding the equilibrium cancellation
proved below. Thus $\mathcal L(U_{\rm rest})=0$ and
$E_\tau(U_{\rm rest})=U_{\rm rest}$: each Euler stage preserves the
lake-at-rest state, as does the two-stage time-stepping scheme below.


With piecewise-constant reconstruction, the preceding construction is a first-order
kinetic hydrostatic finite-volume method. Limited linear reconstructions of
$(\surface,h,\velphi,\veltheta)$ give second-order spatial reconstruction
in smooth resolved wet regions; limiting and the first-order reconstruction on shallow stencils
reduce the local order near nonsmooth or shallow states.

The CFL restriction depends on the current state and may change after the first Euler stage. We therefore allow independently chosen stage steps, so that each stage satisfies its own positivity restriction without restarting the update. For time integration, use two forward-Euler steps with positive step sizes $\Delta t_1$ and $\Delta t_2$:
\begin{equation}\label{eq:ps-variable-rk}
\begin{aligned}
 U^{(1)}&=E_{\Delta t_1}(U^k),\qquad
 U^{(2)}=E_{\Delta t_2}(U^{(1)}),\\
 U^{k+1}&=
 \frac{\Delta t_1^2+\Delta t_2^2}{(\Delta t_1+\Delta t_2)^2}U^k
 +\frac{2\Delta t_1\Delta t_2}{(\Delta t_1+\Delta t_2)^2}U^{(2)},\\
 t^{k+1}&=t^k+\Delta t,\qquad
 \Delta t=\frac{2\Delta t_1\Delta t_2}{\Delta t_1+\Delta t_2}.
\end{aligned}
\end{equation}
This Runge--Kutta update is second order in time for smooth solutions
when the ratio $\Delta t_2/\Delta t_1$ remains bounded above and away from
zero under refinement. Its final weights are nonnegative and sum to one.
For equal stage steps it reduces to the usual SSPRK2 method. The physical
time increment is $\Delta t$, not $\Delta t_1+\Delta t_2$.
At each Euler stage the reconstructed face values, equilibria, interface integrals and sources
are recomputed from that stage's state. Equilibria are reconstructed from
moments, without a separate numerical collision substep.

For the Courant--Friedrichs--Lewy (CFL) restriction, define the maximum
face support speed from the reconstructed state by
\[
 a_f(U)=\max_{s=L,R}\left\{|u_s(e_\perp)|+\sqrt{5\ggra h_s^*/2}\right\}.
\]
Let $\Delta t_{\rm CFL}(U)$ be the largest Euler step satisfying
\begin{equation}\label{eq:ps-stage-cfl}
 \Delta t_{\rm CFL}(U)\max_K\left\{
 \frac{2}{A_K}\sum_{f\subset\partial K}L_fa_f(U),\;
 |f_j|+|\kappa_j|\left(
 \left|\frac{u_{\phi,K}}{R\sin\theta_K}\right|+
 \left|\frac{u_{\theta,K}}R\right|\right)+\gamma_K\right\}
 \leq C_{\rm CFL}.
\end{equation}
Here $0<C_{\rm CFL}\leq1$. The factor two in the transport rate accounts for the bound
$h_{Kf}\leq2h_K$ on reconstructed face depths. We choose
$C_{\rm CFL}=0.8$ as a safety factor. The explicit-source rate is an
additional time-step safeguard, not an energy-stability assertion.
It includes $\gamma_K$ so that $\tau\gamma_K\leq1$ at each stage. Manning friction can make this
explicit restriction severe near dry states; implicit or exact friction
integration would require a separate time-discretization analysis.
Choose a finite $0<\Delta t_1\leq\Delta t_{\rm CFL}(U^k)$, form $U^{(1)}$,
and then choose $0<\Delta t_2\leq\Delta t_{\rm CFL}(U^{(1)})$.
Thus each Euler step satisfies its own restriction without repeating the
first step. Positivity does not require the stage steps to be comparable;
comparability is the separate qualification for the second-order accuracy
statement above. 

\subsection{Properties of the numerical scheme}
\label{sec:ps-properties}

We follow the questions addressed by PS's Theorem~3, but distinguish the properties of the present scheme from those proved there.

\begin{proposition}
\label{prop:ps-water-conservation}
On the full spherical grid with a time-independent round metric and fixed cell areas, periodic longitude boundaries, and zero-length
polar faces, every forward-Euler update \eqref{eq:ps-euler}
preserves the total discrete water volume. The two-stage Runge--Kutta update
\eqref{eq:ps-variable-rk} preserves the same quantity:
\begin{equation}\label{eq:ps-mass-conservation}
 \sum_K A_K h_K^{\rm E}=\sum_K A_Kh_K,
 \qquad \sum_K A_K h_K^{k+1}=\sum_K A_Kh_K^k.
\end{equation}
These identities hold in exact arithmetic, independently of the positive
stage step sizes and without a CFL restriction.
\end{proposition}
\begin{proof}
For a three-component flux $F$, let $F^h$ denote its water-depth component.
The source~\eqref{eq:ps-raw-source} and the pressure corrections have zero
mass component. At every shared face,
$(\widehat F_{Lf})^h=(\widehat F_{Rf})^h=(F_f^*)^h$.
Multiplying the depth component of \eqref{eq:ps-euler} by $A_K$ and summing
over all cells cancels each shared flux, since its face length is the same
on both sides and its outward signs are opposite. Fluxes where the first and last longitude columns meet cancel by periodicity;
the polar faces contribute zero because their lengths vanish. This proves the first identity.
Applying it to both Euler stages gives the same total volume for $U^k$,
$U^{(1)}$, and $U^{(2)}$. The final weights in
\eqref{eq:ps-variable-rk} sum to one, proving the second identity.
\end{proof}
The conserved quantity is area-weighted water volume; multiplication by a
constant material density gives physical mass.

\begin{proposition}
\label{prop:ps-positivity}
Assume a time-independent round metric and the fixed spherical grid above. Assume that the cell depths are nonnegative, the friction rates are finite and nonnegative, velocities are finite and
are set to zero in dry cells, and the reconstruction and hydrostatic cuts
are those specified above. For a forward-Euler step of size $\tau>0$,
the sufficient restriction
\begin{equation}\label{eq:ps-positivity-cfl}
 \frac{\tau}{A_K}\sum_{f\subset\partial K}L_fa_f(U)\leq\frac12
 \qquad\text{for every cell }K
\end{equation}
ensures $h_K^{\rm E}\geq0$. Consequently,
\eqref{eq:ps-variable-rk} preserves nonnegative depth at both Euler stages
and at the completed time level when \eqref{eq:ps-positivity-cfl} holds
with $(U,\tau)=(U^k,\Delta t_1)$ and $(U^{(1)},\Delta t_2)$.
In particular, the sequential choices prescribed by
\eqref{eq:ps-stage-cfl} satisfy these conditions. For $C_{\rm CFL}<1$,
initially positive cell depths remain positive in exact arithmetic.
\end{proposition}
\begin{proof}
The slope restriction $|D_rh_K|\leq2h_K$ and the first-order reconstruction give
$0\leq h_{Kf}\leq2h_K$. Since $B_*\geq B_{Kf}$, hydrostatic reconstruction
therefore gives
\[
 0\leq h_{Kf}^*=\max(h_{Kf}+B_{Kf}-B_*,0)\leq h_{Kf}\leq2h_K.
\]
Let $M_{Kf}^*$ denote the nonnegative equilibrium formed from this cut depth
and the reconstructed velocity of cell $K$ at face $f$. Its total mass is
$h_{Kf}^*$ and its normal particle speed ($\xi_\perp$) is bounded in magnitude by $a_f(U)$.
The outward normal speed is $\sigma_{Kf}\xi_{\perp,f}$, so its outgoing
mass moment satisfies
\[
 0\leq\int_{T_{x_f}^*S_R^2}\max(\sigma_{Kf}\xi_{\perp,f},0)M_{Kf}^*\,\mathrm d\mu_{x_f}
 \leq a_f(U)h_{Kf}^*.
\]
Here $x_f$ is the face evaluation point; in the local orthonormal frame, the integral ranges over $(\xi_\perp,\xi_\parallel)\in\mathbb R^2$.
The upwind flux supplies a nonnegative incoming mass contribution from the
neighboring cell. Pressure corrections and all source terms have zero mass
component. Dropping the incoming contributions from the Euler depth update
thus yields
\begin{equation}\label{eq:ps-depth-bound}
\begin{aligned}
 h_K^{\rm E}
 &\geq h_K-\frac{\tau}{A_K}
       \sum_{f\subset\partial K}L_fa_f(U)h_{Kf}^*
 \geq h_K\left(1-\frac{2\tau}{A_K}
       \sum_{f\subset\partial K}L_fa_f(U)\right)\geq0.
\end{aligned}
\end{equation}
This also covers $h_K=0$, for which every outgoing cut depth vanishes.
Under \eqref{eq:ps-stage-cfl}, the last bound is at least
$(1-C_{\rm CFL})h_K$, giving strict positivity for initially wet cells
when $C_{\rm CFL}<1$. Applying this argument successively to the two Euler
steps and then taking the nonnegative convex combination in
\eqref{eq:ps-variable-rk} proves the assertion.
\end{proof}

\begin{proposition}
\label{prop:ps-lake-at-rest}
Assume a time-independent round metric, a fixed spherical grid, and time-independent bottom topography. For a constant level $\ell$, consider the discrete state
\[
 \frac{u_{\phi,K}}{R\sin\theta_K}
 =\frac{u_{\theta,K}}{R}=0,
 \qquad h_K=\max(\ell-B_K,0).
\]
With the reconstruction and dry-stencil rule specified above, the scheme
is well balanced for this state, including wet/dry interfaces:
$\mathcal L(U)_K=0$ in every cell. Consequently, both Euler stages and
the Runge--Kutta update \eqref{eq:ps-variable-rk} preserve the state
exactly in exact arithmetic.
\end{proposition}

\begin{proof}
On wet stencils $D_r\surface_K=0$; the dry-stencil rule sets it to zero
on the remaining stencils relevant to the wet/dry interface. A dry cell
has zero reconstructed depth at every face. At a face between two wet cells the two cut depths in
\eqref{eq:ps-interface-depths} coincide; at a blocked wet/dry face both
are zero. Write $h_f^*=h_L^*=h_R^*$ for this common cut depth. Since $T(h,0)$ agrees on the two sides and $P(h,0)=p(h)/2$,
\[
 (F_f^*)^h=0,\qquad F_f^*=p(h_f^*)\mathbf b_r,
 \qquad \mathcal R_{Kf}=0.
\]
Bottom friction also vanishes because the momentum is zero. All terms in the residual source \eqref{eq:ps-residual-source} vanish.
Consequently $E_\tau(U)=U$ for either stage step, and the convex update
\eqref{eq:ps-variable-rk} also leaves this state unchanged, up to
roundoff in floating-point arithmetic.
\end{proof}

\section{Numerical results}
\label{sec:numerical-results}
\newcommand{\EarthInitialPeak}{4.9251}
\newcommand{\EarthKineticSteps}{144192}
\newcommand{\EarthKineticMin}{0.0004248}
\newcommand{\EarthKineticMass}{$\num{9.43e-13}$}
\newcommand{\EarthRatioMin}{1.000000}
\newcommand{\EarthRatioMax}{1.000000}
\newcommand{\EarthRatioDeparture}{$\num{3.89e-15}$}

\subsection{Global propagation over Earth topography}
\label{sec:earth-kinetic}
We consider a localized free-surface displacement on the rotating Earth,
using the spherical kinetic method of the preceding section. This is an
idealized long-wave propagation experiment, rather than a reconstruction of
a particular earthquake. The domain is the entire sphere with its fixed round metric and parameters
$R=6\,371\,000\ \mathrm{m}$, $\ggra=9.81\ \mathrm{m\,s^{-2}}$, and
$\varpi=7.292115\times10^{-5}\ \mathrm{s^{-1}}$; hence
$f(\theta)=2\varpi\cos\theta$. The bottom topography varies spatially but is
independent of time; atmospheric forcing and wave dispersion are not included.

The main experiment uses $720\times360$ longitude--colatitude cells, so that
$\Delta\phi=\Delta\theta=0.5^\circ$. The north--south cell width is about
$55.6\ \mathrm{km}$, while the east--west width decreases with latitude.
We use the exact spherical cell areas and face lengths specified above,
two ghost layers, periodic continuation in longitude, and antipodal
ghost-cell continuation at the poles, where the polar faces have zero length.
Bottom elevations $B_K$ are obtained by bilinear interpolation to cell
centers of the locally stored GMT gridline-registered one-degree relief dataset
\code{earth_relief_01d_g.grd} \cite{gmtEarthRelief}. Its stored metadata
identify SRTM15+ version 2.4 as the source, with a Gaussian filter of
$111.2\ \mathrm{km}$ full width. Tozer et al.\ \cite{tozer2019srtm}
describe the underlying SRTM15+ dataset. The same interpolated
bottom is used for initialization, evolution, and coastline visualization;
the sea level is zero and no vertical shift is applied. This coarse,
filtered relief resolves basin-scale propagation, not coastal inundation
or fine bathymetric focusing.

The kinetic method and the central-upwind method described in
Section~\ref{sec:earth-comparison} both include Manning bottom friction with coefficient $n_{\mathrm M}=0.025\,\mathrm{s\,m^{-1/3}}$, a generic tsunami-modeling
choice~\cite{geoclawManning}, rather than a calibration to local seabed roughness.
To avoid stiffness caused by the artificial millimeter-depth land film,
only the friction denominator is regularized below $10^{-2}\,\mathrm m$:
\[
 \gamma_K=\frac{g_{\mathrm{grav}}n_{\mathrm M}^2
       \sqrt{q_{E,K}^2+q_{S,K}^2}}
       {\max(h_K,10^{-2}\,\mathrm m)^{7/3}}.
\]
This is the Manning rate for $h_K\geq10^{-2}\,\mathrm m$ with a regularized denominator below that threshold, applied identically in both methods.
It changes neither water depth nor velocity recovery; the source remains
opposed to momentum and vanishes at rest. Each Runge--Kutta stage recomputes
this rate from the true cell-mean momenta, includes it in the explicit
source-step restriction, and adds $-\gamma_K(q_{E,K},q_{S,K})$ to the momentum
source. The transport reconstruction and its near-dry treatment remain
specific to each method.

Let $d_K$ be the angular great-circle distance from the center of cell $K$
to the point of latitude $15^\circ$ N and longitude $150^\circ$ W. This is the angle subtended at the sphere's center by the shortest surface arc joining the two points; its physical length is $Rd_K$, with $d_K$ in radians. With
$\lambda_0=15^\circ$ and $\phi_0=-150^\circ$, it satisfies
\[
 \cos d_K=\sin\lambda_0\cos\theta_K
 +\cos\lambda_0\sin\theta_K\cos(\phi_K-\phi_0).
\]
The initial physical velocities vanish, and the depth is
\begin{equation}\label{eq:earth-initial-data}
 h_K(0)=\max\{-B_K,\varepsilon\}
 +a_0\exp\!\left(-\frac{d_K^2}{2\sigma^2}\right)
       \boldsymbol{1}_{\{B_K<0\}},
 \qquad
 a_0=5\ \mathrm{m},\quad \sigma=2^\circ,\quad
 \varepsilon=10^{-3}\ \mathrm{m}.
\end{equation}
Angles are converted to radians when evaluating this formula. These
cell-center samples initialize the stored finite-volume states. The
Gaussian is confined to the ocean and has physical width $R\sigma\simeq
222.4\ \mathrm{km}$. Its sampled maximum is \EarthInitialPeak\ m, slightly
below the prescribed amplitude because its center lies between cell
centers. The positive land film regularizes the initial data but is not a
motionless equilibrium over sloping land: its free surface is $B+\varepsilon$.
Consequently, this experiment is not an exact lake-at-rest test.
The positive pulse adds water relative to the reference depth; volume
conservation is measured relative to this prescribed perturbed initial state.

We evolve the solution for six hours with minmod parameter $\beta=1.5$,
the shallow-stencil threshold $h_{\rm rec}=10^{-2}\ \mathrm{m}$, and
$C_{\rm CFL}=0.8$ in \eqref{eq:ps-stage-cfl}. Each of the two Euler stages
uses its own admissible step, and their convex combination advances time
by equation \eqref{eq:ps-variable-rk}. The kinetic computation preserves positive depths using the stagewise CFL restriction.
The run takes \EarthKineticSteps\ time steps; its minimum
depth over all Euler stages and completed time-step states is \EarthKineticMin\ m.
The maximum relative water-volume change at the saved times is
\EarthKineticMass. The largest recorded departure of $\Delta t_2/\Delta t_1$ from unity
is \EarthRatioDeparture. Thus the stage steps happen to be nearly equal
in this experiment; it does not exercise substantially unequal stage steps.

\begin{figure}[p]
 \centering
 \includegraphics[width=0.81\textwidth,height=0.78\textheight,keepaspectratio]{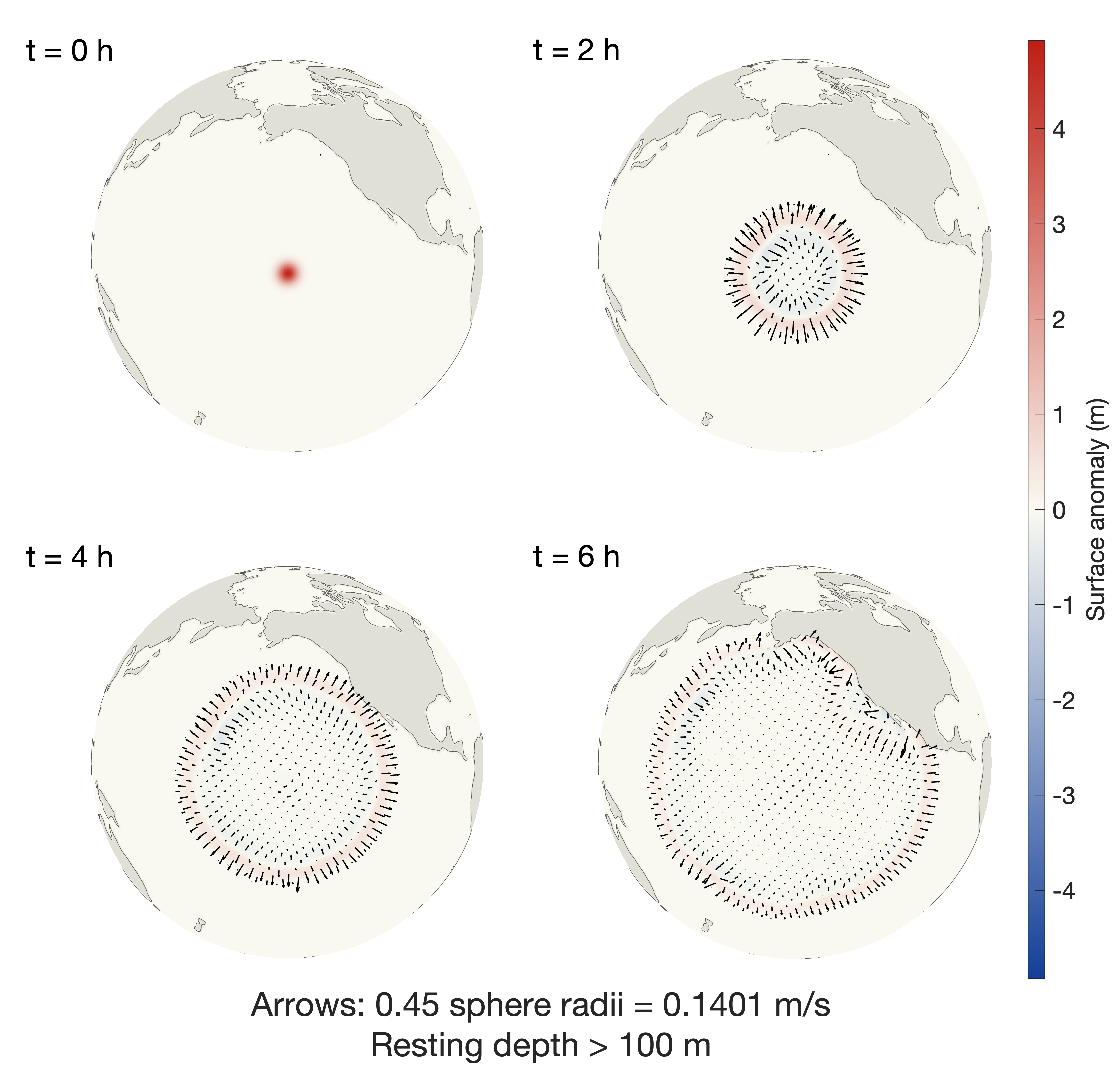}
 \caption{Kinetic evolution of the localized Pacific displacement on the
 $720\times360$ spherical mesh with a prescribed $5\,\mathrm m$ initial amplitude and Manning bottom friction. Panels show $t=0,2,4,6$ hours in reading
 order, each centered at longitude $150^\circ$ W and latitude $20^\circ$ N. Colors represent $\delta\surface$ in metres,
 and arrows represent $u^\sharp=(q_E/h)e_\phi+(q_S/h)e_\theta$ in $\mathrm{m\,s^{-1}}$, with the common scale shown in the figure. The land-film region is masked in the ocean
 color and arrow diagnostics.}
 \label{fig:earth-kinetic-snapshots}
\end{figure}

For visualization and comparison we subtract the common reference depth,
\[
 \delta\surface_K(t)=h_K(t)-\max\{-B_K,\varepsilon\}.
\]
In ocean cells deeper than the film this equals the free-surface elevation
relative to sea level. Figure~\ref{fig:earth-kinetic-snapshots} shows the Pacific-facing
hemisphere at $0$, $2$, $4$, and $6$ hours. The color and velocity
scales are fixed across times. Land is shown in a neutral color, and
velocity arrows are restricted to cells with resting ocean depth greater
than $100\ \mathrm{m}$, so that land-film velocities do not set their scale.
The arrows plot the physical tangent velocity
\[
 u^\sharp=\frac{q_E}{h}e_\phi+\frac{q_S}{h}e_\theta
 =\frac{u_\phi}{R\sin\theta}e_\phi+\frac{u_\theta}{R}e_\theta,
\]
whose eastward and southward components are $q_E/h$ and $q_S/h$, respectively, in $\mathrm{m\,s^{-1}}$. They are sampled at approximately equally spaced locations on the sphere, with a common physical scale across panels.

The expanding disturbance illustrates propagation on a curved, rotating
surface with spatially varying depth. The closed sphere also avoids imposing an artificial
outer boundary on the propagation problem. For a small surface displacement superposed on
kilometre-scale depths, the pressure--topography cancellation is a useful
feature: the residual source depends on differences of $\surface$, instead
of requiring the numerical subtraction of two large forces. The preserved
water volume and positive stage depths support the conservative and
positivity-preserving behavior of this implementation. 

\subsection{Comparison with a spherical central-upwind method}
\label{sec:earth-comparison}
We compare the kinetic method with a spherical central-upwind solver based on
Kurganov and Petrova~\cite{kurganovPetrova2007}, using identical physical
parameters, grid, topography, initial data, and output times. The comparison
solver uses hydrostatic reconstruction, the same geometric and Coriolis
sources, face speeds $u(e_\perp)\pm\sqrt{\ggra h}$, minmod parameter $1.5$,
SSPRK$(3,3)$, and velocity desingularization below $h=10^{-3}\,\mathrm m$.
Each Euler stage enforces $\Delta t\sum_fL_fa_f/A_K\leq0.45<1/2$,
where $a_f$ is the maximum absolute central-upwind face speed; nonnegative
reconstructed depths and the convex SSPRK combinations preserve nonnegative
depth. The comparison therefore concerns complete methods with different
fluxes, time integrators, CFL restrictions, and near-dry treatments.

\begin{figure}[!htb]
 \centering
 \includegraphics[width=0.82\textwidth]{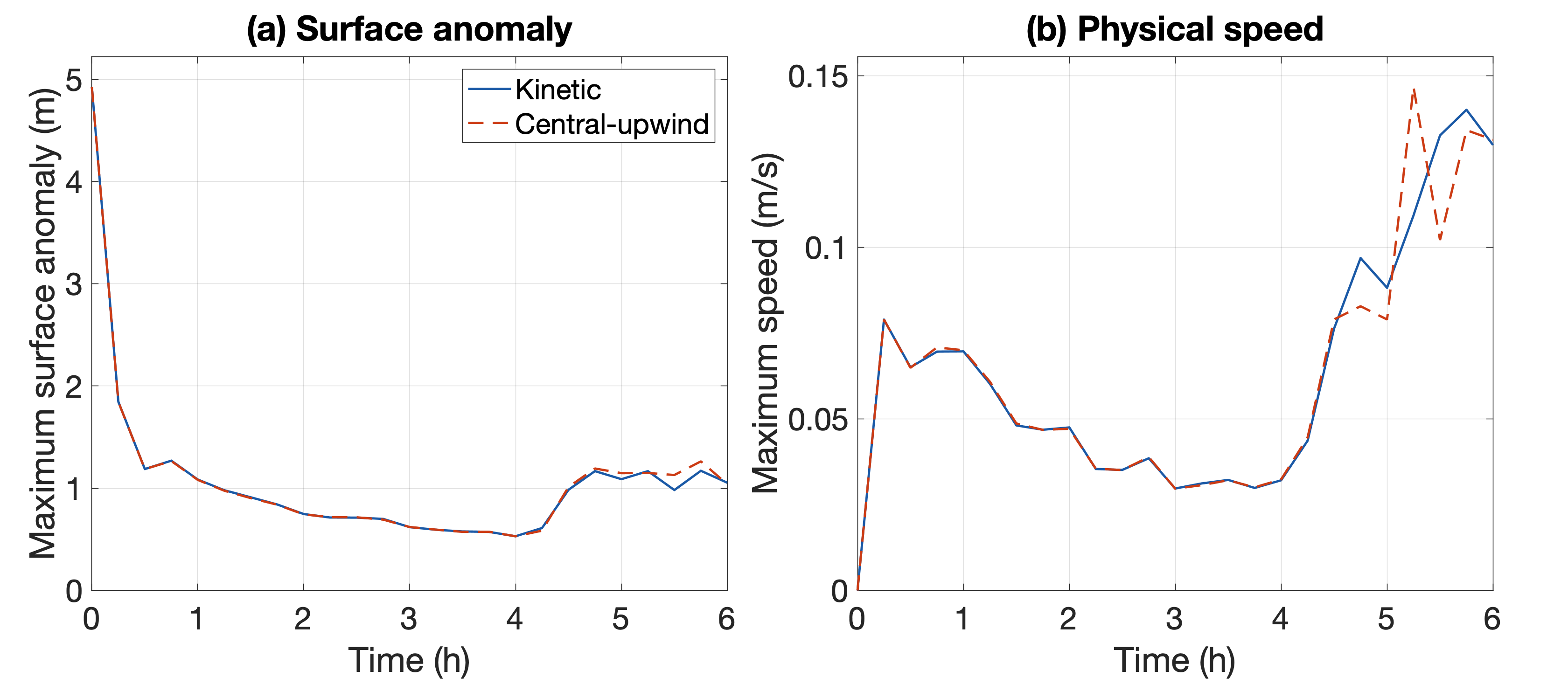}
 \par\smallskip
 {\small\sffamily\bfseries (c) Offshore surface anomaly}\par
 \includegraphics[width=0.78\textwidth]{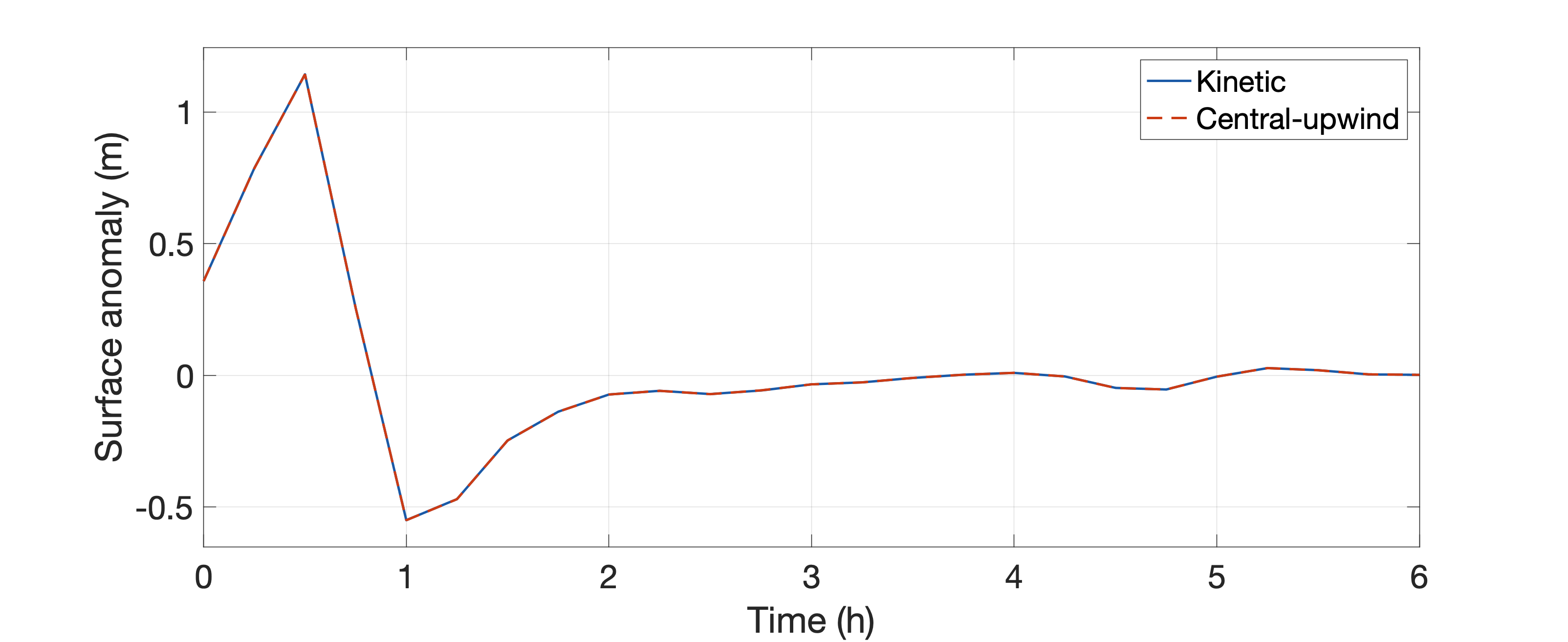}
 \caption{Six-hour comparison on the $720\times360$ mesh with $a_0=5\,\mathrm m$
 and Manning friction: (a) maximum surface anomaly and (b) maximum physical
 speed over cells with resting depth greater than $100\,\mathrm m$;
 (c) surface anomaly at $15.25^\circ$ N, $145.25^\circ$ W
 (resting depth $5395.16\,\mathrm m$). Solid blue and dashed orange curves
 show the kinetic and central-upwind methods, respectively, sampled every
 15 minutes.}
 \label{fig:earth-comparison}
 \label{fig:earth-gauges}
\end{figure}

Figure~\ref{fig:earth-comparison} compares the maximum surface anomaly
$\delta\surface$ (top left), maximum physical speed
$|u|_{g^{-1}}=\sqrt{(q_E/h)^2+(q_S/h)^2}$ (top right), and an offshore surface-anomaly
time series (bottom). The maxima exclude the land film by using the fixed set
$D=\{K:-B_K>100\,\mathrm m\}$. Table~\ref{tab:earth-comparison} also reports
the area-weighted RMS difference
\begin{equation}\label{eq:earth-rms-difference}
 \left(
 \frac{\sum_{K\in D}A_K
  \left(\delta\surface_K^{\mathrm{kin}}-
        \delta\surface_K^{\mathrm{cu}}\right)^2}
 {\sum_{K\in D}A_K}
 \right)^{1/2},
\end{equation}
where the superscripts denote the two numerical solutions. This measures
inter-method differences, not error against an exact solution.

\begin{table}[htbp]
 \centering
 \caption{Ocean diagnostics at the times shown in
 Figure~\ref{fig:earth-kinetic-snapshots}. The two peak columns give
 $\max_{K\in D}\delta\surface_K$; RMS denotes
 \eqref{eq:earth-rms-difference}.}
 \label{tab:earth-comparison}
 \small
 \begin{tabular}{rrrrrr}
\toprule
Time (h) & \multicolumn{2}{c}{Peak anomaly (m)} & RMS (m) & \multicolumn{2}{c}{Max. speed (m/s)}\\
 & Kinetic & Central-upwind & & Kinetic & Central-upwind\\
\midrule
0 & 4.92510 & 4.92510 & 0.000000 & 0.00000 & 0.00000\\
2 & 0.74744 & 0.74412 & 0.000281 & 0.04752 & 0.04714\\
4 & 0.53017 & 0.52994 & 0.000341 & 0.03212 & 0.03239\\
6 & 1.05359 & 1.03869 & 0.005033 & 0.12985 & 0.13139\\
\bottomrule
\end{tabular}

\end{table}

The methods agree closely in offshore propagation
(Figure~\ref{fig:earth-comparison} and Table~\ref{tab:earth-comparison}).
At six hours, their maximum surface anomalies differ by approximately
$1.4\%$ and maximum speeds by $1.2\%$, although larger differences occur
earlier and near continental margins. The area-weighted RMS difference is
$5.03\times10^{-3}\,\mathrm m$, while the largest pointwise difference over
$D$ is $0.40709\,\mathrm m$; the large weakly disturbed ocean area reduces
the RMS. At the observation point in panel (c), the maximum difference is
$0.683\,\mathrm{mm}$, and both sampled peaks occur at $0.5$ hours.
Two additional offshore observation points have maximum differences below
$2.5\,\mathrm{mm}$ and matching peak times at the 15-minute sampling resolution.

The central-upwind run takes $81\,072$ time steps, with maximum relative
water-volume change $3.57\times10^{-14}$ and minimum saved depth
$2.71\times10^{-4}\,\mathrm m$. These results show positive depths and water-volume conservation near
floating-point precision in both runs.

The results may be sensitive to spatial resolution, particularly near
continental margins, possibly because of the wet/dry treatment near the
model coastline, including the land film and near-dry regularization.
Differences in flux dissipation and time integration may also contribute.
Further refinement and comparisons with matched near-dry treatments are
needed to isolate these effects; inter-method agreement alone establishes
neither an accuracy ranking nor validation of coastal inundation.



\section{Conclusions}
\label{sec:conclusions}

The cotangent-bundle formulation provides a common geometric framework for kinetic shallow-water closures on a fixed manifold with a prescribed, possibly time-dependent Riemannian metric. Hamiltonian transport, intrinsic momentum integration, and tensor moment conditions recover the balance laws with hydrostatic pressure, topography, skew-adjoint forcing, and dissipative bottom friction. The explicit Gibbs equilibrium identifies the dimension-dependent relation between scalar and tensor second moments. The resulting equations include the local rate of change of geometric volume, retain the instantaneous shallow-water characteristic speeds, and admit a mechanical-energy balance with geometric work and nonnegative frictional dissipation. Under the stated vanishing-viscosity assumptions, including convergence of geometric work, the corresponding entropy inequality holds in the weak sense. Total mass is conserved with the evolving volume element; the energy monotonicity and stationary lake results are stated for the fixed-metric specialization. Specialization to the rotating sphere with its fixed round metric connects this intrinsic description to the familiar physical velocity components and geometric source terms.

The spherical finite-volume method combines kinetic fluxes with hydrostatic reconstruction and a consistent treatment of geometry and forces. It conserves discrete water volume, is well balanced for lakes at rest including wet/dry interfaces, and is positivity preserving under the derived CFL restriction. The two-stage second-order time integrator preserves nonnegative depth by enforcing the CFL restriction. These properties follow from the discrete flux and source structure; a discrete entropy inequality for the reconstructed scheme remains to be established.

The refined Earth-topography experiment preserves positive stage depths and water volume, with maximum relative volume change \EarthKineticMass\ for the kinetic method. Both methods give similar offshore gauge time series. At six hours their surface and speed maxima differ by approximately $1.4\%$ and $1.2\%$, respectively, although larger differences occur at earlier times and locally near continental margins. Inter-method agreement alone does not establish grid convergence or an accuracy ranking. 

Further work should assess convergence, wetting and drying, and sensitivity to Manning roughness and its near-dry regularization. Establishing a discrete entropy inequality while retaining well-balancing, positivity, and second-order accuracy remains open. Applications to other surface geometries require gravity and additional forces appropriate to the physical setting. Europa provides a motivation for prescribed evolving metrics, but coupling the fluid to an ice shell and tidal forcing remains future work. A numerical method for an evolving metric would also need to account for changing cell areas and geometric work; the present spherical scheme and experiments address only a fixed metric.


\appendix
\section{Alternative numerical scheme via semiclassical quantization}
\label{app:semiclassical}


This appendix describes an exploratory semiclassical approximation, separate from the finite-volume method analyzed above. On the unit circle with its fixed standard metric, no rotation force, and $\gamma=0$, equation \eqref{eq:sw-kinetic} becomes 

\begin{equation}
\label{eq:QuantumKinetic}
\dpar{M}{t}+\left\{ M(t,x,\xi), H(x,\xi) \right\} =C(t,x,\xi),
\end{equation}
Use the periodic coordinate $s\in\mathbb R/(2\pi\mathbb Z)$, with $x=e^{is}$ and $\xi=\xi_s\,ds$. For the unit-circle metric $g=ds\otimes ds$, the Hamiltonian is
$
H(s,\xi_s)=\frac{\xi_s^2}{2}+g_{\mathrm{grav}}B(s). \; 
$

$M$ is a phase-space symbol. Denote its Weyl quantization by $M^W$ and use $\hbar>0$ as a numerical semiclassical parameter, rather than a physical Planck constant.

The equilibrium has a square-root singularity at its support boundary. Smooth-symbol semiclassical calculus therefore does not apply directly there. The singular-symbol calculus of \cite{hernandez2015algebras} provides a related framework for Zoll-type domains. 

The formal operator analogue is
\[
\dpar{M^W}{t}+\frac{i}{\hbar} \left[ H^W,M^W \right]=C^W,
\]
where the commutator is defined by
\[
 [H^W,M^W]=H^WM^W-M^WH^W.
\]
For smooth symbols, the principal symbol of $(i/\hbar)[H^W,M^W]$ is $\{M,H\}$ with our bracket convention. Thus this is a semiclassical correspondence, not an exact identification of the kinetic and operator equations. On the circle,
\[
 H^W=-\frac{\hbar^2}{2}\frac{\partial^2}{\partial s^2}
       +g_{\mathrm{grav}}B(s).
\]

For a collisionless transport substep with fixed $B$, the operator equation has the exact solution
\begin{equation}\label{eq:SolMat}
 M^W(t)=e^{-itH^W/\hbar}M^W(0)e^{itH^W/\hbar}.
\end{equation}
Vanishing collision moments do not mean that the collision term itself vanishes. A shallow-water kinetic algorithm must also restore the equilibrium moment closure between transport substeps. Egorov theory describes the smooth-symbol semiclassical transport for fixed times \cite{zworski}. 

The cotangent bundle of the circle is a cylinder. For a finite momentum interval $[\xi_0,\xi_1]$ and $\hbar=1/N$, $N\in\mathbb N$, define the orthogonal Fourier projection
\[
 \pi_N:L^2(S^1)\longrightarrow\mathcal H_N,
 \qquad
 \mathcal H_N=\operatorname{span}\{e^{iks}:k\in\mathbb Z,\ \xi_0\leq k/N\leq\xi_1\}.
\]
The parameter $N$ sets the momentum spacing $1/N$.  The distribution is $2\pi$-periodic in $s$, not in $\xi_s$. We write
\[
 M(t,s,\xi_s)=\sum_{\ell\in\mathbb Z}a_\ell(t,\xi_s)e^{i\ell s}.
\]
In the normalized Fourier basis, the compressed Weyl matrix is
\[
 (M_N)_{jk}=a_{j-k}\!\left(t,\frac{j+k}{2N}\right),
 \qquad j,k\in\mathbb Z\cap[N\xi_0,N\xi_1].
\]
Choose the interval to contain the relevant momentum support. For a finite-dimensional transport step, set $H_N=\pi_NH^W\pi_N|_{\mathcal H_N}$ and evolve
\[
 M_N(t)=e^{-itH_N/\hbar}M_N(0)e^{itH_N/\hbar}.
\]
This unitary conjugation preserves the trace and, if present initially, positive semidefiniteness. Weyl quantization of a nonnegative phase-space symbol need not yield a positive semidefinite matrix; this property is therefore an additional condition, not a consequence of $M\geq0$. It is distinct from compressing the full propagator unless the subspace is invariant under $H^W$. Summation along matrix diagonals, with momentum-spacing weights $1/N$ and the corresponding momentum factor for $hu$, approximates the Fourier coefficients of the momentum integrals defining $h$ and $hu$. This is how we recover the macroscopic quantities. Trace invariance gives conservation of the reconstructed mean depth during this transport substep. Nonnegative reconstructed depth follows if the matrix is positive semidefinite and the diagonal sums are evaluated exactly; it is not an unconditional positivity result for this exploratory method or its equilibrium reconstruction. 



\begin{figure}[htbp]
\begin{center}
\resizebox{6.0in}{3.0in}
{\includegraphics[width=4.0cm]{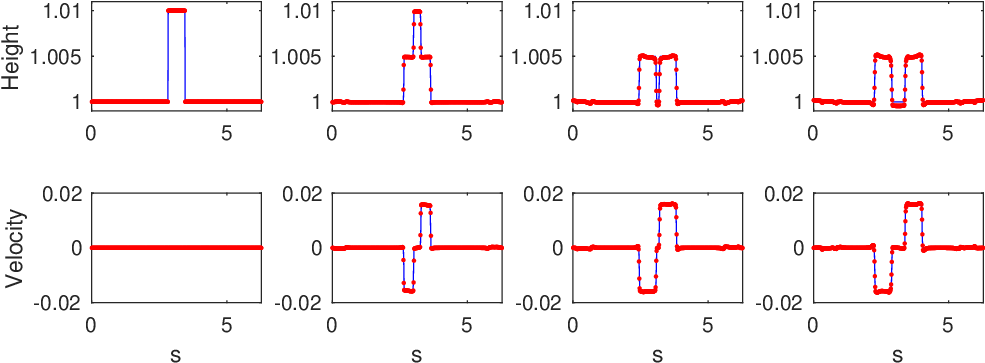}}
\end{center}
\caption{\label{fig:UpwindQuantum} Illustrative circle comparison: depth (top) and velocity (bottom) from the second-order upwind method (solid blue) and the semiclassical method (red dotted line), at nominal output times $t=0,0.06,0.12,0.18$. }
\end{figure} 

The illustrative circle experiment uses a flat bottom $B=0$, $g_{\mathrm{grav}}=9.81$, zero initial velocity, and the periodic initial depth
\[
 h(0,s)=1+10^{-2}\mathbf1_{(0.45(2\pi),\,0.55(2\pi))}(s).
\]
The stored implementation uses $N=500$ spatial samples and momentum interval $[-10,10]$. Figure \ref{fig:UpwindQuantum} shows the water depth (top panels) and velocity (bottom panels) at nominal times $t=0,0.06,0.12,0.18$ using the second-order upwind scheme (blue solid line) and the quantum scheme (red dotted line). The two schemes show good agreement in both water depth and velocity at the displayed times.

Extending this spectral construction from the circle to the sphere would
replace the Fourier modes by spherical harmonics, the eigenfunctions of
the spherical Laplace--Beltrami operator. Developing and analyzing this
extension, including the treatment of rotation and bottom friction, is
left for future work.

\section*{Acknowledgements}
This research was supported in part by UNAM-DGAPA-PAPIIT grant IN115925.

\section*{Disclosure of AI assistance}
OpenAI Codex was used to assist with manuscript revision and the development and verification of the numerical code. The author retains full responsibility for the mathematical arguments, numerical results, and final content of the manuscript.

\bibliographystyle{unsrt}
\bibliography{References}

@article{Roe1981,
  author  = {Roe, Philip L.},
  title   = {Approximate Riemann Solvers, Parameter Vectors,
             and Difference Schemes},
  journal = {Journal of Computational Physics},
  volume  = {43},
  number  = {2},
  pages   = {357--372},
  year    = {1981},
  doi     = {10.1016/0021-9991(81)90128-5}
}

@inproceedings{KastFidkowskiRoe2011,
  author    = {Kast, Steven M. and Fidkowski, Krzysztof J.
               and Roe, Philip L.},
  title     = {An Unsteady Entropy Adjoint Approach for
               Adaptive Solution of the Shallow-Water Equations},
  booktitle = {20th AIAA Computational Fluid Dynamics Conference},
  year      = {2011},
  publisher = {American Institute of Aeronautics and Astronautics},
  doi       = {10.2514/6.2011-3694}
}

@article{VasconcelosWrightRoe2006a,
  author  = {Vasconcelos, Jose G. and Wright, Steven J.
             and Roe, Philip L.},
  title   = {Improved Simulation of Flow Regime Transition
             in Sewers: Two-Component Pressure Approach},
  journal = {Journal of Hydraulic Engineering},
  volume  = {132},
  number  = {6},
  pages   = {553--562},
  year    = {2006},
  doi     = {10.1061/(ASCE)0733-9429(2006)132:6(553)}
}

@article{VasconcelosWrightRoe2006b,
  author  = {Vasconcelos, Jose G. and
             Wright, Steven J. and
             Roe, Philip L.},
  title   = {Current Issues on Modeling Extreme
             Inflows in Stormwater Systems},
  journal = {Journal of Water Management Modeling},
  year    = {2006},
  pages   = {R225-19},
  doi     = {10.14796/JWMM.R225-19}
}

@article{VasconcelosWrightRoe2009,
  author  = {Vasconcelos, Jose G. and Wright, Steven J.
             and Roe, Philip L.},
  title   = {Numerical Oscillations in Pipe-Filling Bore
             Predictions by Shock-Capturing Models},
  journal = {Journal of Hydraulic Engineering},
  volume  = {135},
  number  = {4},
  pages   = {296--305},
  year    = {2009},
  doi     = {10.1061/(ASCE)0733-9429(2009)135:4(296)}
}

@book{zworski,
  title={Semiclassical analysis},
  author={Zworski, Maciej},
  series={Graduate Studies in Mathematics},
  volume={138},
  year={2012},
  publisher={American Mathematical Soc.}
}

@article{hernandez2015algebras,
  title={Algebras of semiclassical pseudodifferential operators associated with Zoll-type domains in cotangent bundles},
  author={Hernandez-Duenas, Gerardo and Uribe, Alejandro},
  journal={Journal of Functional Analysis},
  volume={268},
  number={7},
  pages={1755--1807},
  year={2015},
  publisher={Elsevier},
  doi={10.1016/j.jfa.2014.12.004},
  url={https://doi.org/10.1016/j.jfa.2014.12.004},
  note={\href{https://doi.org/10.1016/j.jfa.2014.12.004}{doi:10.1016/j.jfa.2014.12.004}}
}

@article{mueller1992,
  author = {M\"uller, D.},
  title = {{Angular-momentum budget of shallow waters on a rotating sphere}},
  journal = {Physical Review A},
  volume = {45},
  year = {1992},
  pages = {5545--5555},
  doi = {10.1103/PhysRevA.45.5545},
  url = {https://doi.org/10.1103/PhysRevA.45.5545},
  note = {\href{https://doi.org/10.1103/PhysRevA.45.5545}{doi:10.1103/PhysRevA.45.5545}}
}

@article{arpaia2022,
  author = {Arpaia, L. and Ricchiuto, M. and Filippini, A. G. and Pedreros, R.},
  title = {{An efficient covariant frame for the spherical shallow water equations: Well balanced DG approximation and application to tsunami and storm surge}},
  journal = {Ocean Modelling},
  volume = {169},
  year = {2022},
  pages = {101915},
  doi = {10.1016/j.ocemod.2021.101915},
  url = {https://doi.org/10.1016/j.ocemod.2021.101915},
  note = {\href{https://doi.org/10.1016/j.ocemod.2021.101915}{doi:10.1016/j.ocemod.2021.101915}}
}

@article{montoya2026,
  author = {Montoya, T. and Rueda-Ram\'irez, A. M. and Gassner, G. J.},
  title = {{Entropy-stable discontinuous spectral-element methods for the spherical shallow water equations in covariant form}},
  journal = {Journal of Computational Physics},
  volume = {555},
  year = {2026},
  pages = {114782},
  doi = {10.1016/j.jcp.2026.114782},
  url = {https://doi.org/10.1016/j.jcp.2026.114782},
  note = {\href{https://doi.org/10.1016/j.jcp.2026.114782}{doi:10.1016/j.jcp.2026.114782}. Open manuscript: \href{https://arxiv.org/html/2509.08790v2}{arXiv:2509.08790v2}}
}

@article{perthame2001,
  author = {Perthame, B. and Simeoni, C.},
  title = {{A kinetic scheme for the Saint-Venant system with a source term}},
  journal = {Calcolo},
  volume = {38},
  year = {2001},
  pages = {201--231},
  doi = {10.1007/s10092-001-8181-3},
  url = {https://doi.org/10.1007/s10092-001-8181-3},
  note = {\href{https://doi.org/10.1007/s10092-001-8181-3}{doi:10.1007/s10092-001-8181-3}}
}

@article{mangeney2003,
  author = {Mangeney-Castelnau, A. and Villotte, J.-P. and Bristeau, M.-O. and Perthame, B. and Bouchut, F. and Simeoni, C. and Yerneni, S.},
  title = {{Numerical modeling of avalanches based on Saint Venant equations using a kinetic scheme}},
  journal = {Journal of Geophysical Research: Solid Earth},
  volume = {108},
  number = {B11},
  year = {2003},
  pages = {2527},
  doi = {10.1029/2002JB002024},
  url = {https://doi.org/10.1029/2002JB002024},
  note = {\href{https://doi.org/10.1029/2002JB002024}{doi:10.1029/2002JB002024}}
}

@article{audusse2011,
  author = {Audusse, E. and Bristeau, M.-O. and Perthame, B. and Sainte-Marie, J.},
  title = {{A multilayer Saint-Venant system with mass exchanges for shallow water flows. Derivation and numerical validation}},
  journal = {ESAIM: Mathematical Modelling and Numerical Analysis},
  volume = {45},
  year = {2011},
  pages = {169--200},
  doi = {10.1051/m2an/2010036},
  url = {https://doi.org/10.1051/m2an/2010036},
  note = {\href{https://doi.org/10.1051/m2an/2010036}{doi:10.1051/m2an/2010036}}
}

@article{audusse2016,
  author = {Audusse, E. and Bouchut, F. and Bristeau, M.-O. and Sainte-Marie, J.},
  title = {{Kinetic entropy inequality and hydrostatic reconstruction scheme for the Saint-Venant system}},
  journal = {Mathematics of Computation},
  volume = {85},
  year = {2016},
  pages = {2815--2837},
  doi = {10.1090/mcom/3099},
  url = {https://doi.org/10.1090/mcom/3099},
  note = {\href{https://doi.org/10.1090/mcom/3099}{doi:10.1090/mcom/3099}}
}

@article{love2012,
  author = {Love, P. J. and Cianci, D.},
  title = {{From the Boltzmann equation to fluid mechanics on a manifold}},
  journal = {Philosophical Transactions of the Royal Society A},
  volume = {369},
  year = {2011},
  pages = {2362--2370},
  doi = {10.1098/rsta.2011.0097},
  url = {https://doi.org/10.1098/rsta.2011.0097},
  note = {\href{https://doi.org/10.1098/rsta.2011.0097}{doi:10.1098/rsta.2011.0097}. Open manuscript: \href{https://arxiv.org/abs/1208.5514}{arXiv:1208.5514}}
}

@article{gonzalez2026,
  author = {Gonz\'alez, A. and Castro, M. J. and Mac\'ias, J.},
  title = {{Asymptotically well-balanced geostrophic reconstruction finite volume numerical schemes for the two-dimensional rotating NLSWE in spherical coordinates}},
  journal = {SIAM Journal on Applied Mathematics},
  volume = {86},
  number = {4},
  year = {2026},
  pages = {1974--2002},
  doi = {10.1137/25M1726522},
  url = {https://doi.org/10.1137/25M1726522},
  note = {\href{https://doi.org/10.1137/25M1726522}{doi:10.1137/25M1726522}}
}

@article{audusse2005,
  author = {Audusse, E. and Bristeau, M.-O.},
  title = {A well-balanced positivity preserving ``second-order'' scheme for shallow water flows on unstructured meshes},
  journal = {Journal of Computational Physics},
  volume = {206},
  number = {1},
  year = {2005},
  pages = {311--333},
  doi = {10.1016/j.jcp.2004.12.016},
  url = {https://doi.org/10.1016/j.jcp.2004.12.016},
  note = {\href{https://doi.org/10.1016/j.jcp.2004.12.016}{doi:10.1016/j.jcp.2004.12.016}}
}

@article{staniforth2015,
  author = {Staniforth, Andrew and White, Andy},
  title = {The shallow-water equations in non-spherical geometry with latitudinal variation of gravity},
  journal = {Quarterly Journal of the Royal Meteorological Society},
  volume = {141},
  number = {687},
  pages = {655--662},
  year = {2015},
  doi = {10.1002/qj.2394},
  url = {https://doi.org/10.1002/qj.2394}
}

@article{bouchut2004topography,
  author = {Bouchut, Fran\c{c}ois and Westdickenberg, Michael},
  title = {Gravity driven shallow water models for arbitrary topography},
  journal = {Communications in Mathematical Sciences},
  volume = {2},
  number = {3},
  pages = {359--389},
  year = {2004},
  doi = {10.4310/CMS.2004.v2.n3.a2},
  url = {https://doi.org/10.4310/CMS.2004.v2.n3.a2}
}

@misc{gmtEarthRelief,
  author = {{GMT Team}},
  title = {{IGPP Earth Relief}: Remote Datasets},
  howpublished = {\url{https://www.generic-mapping-tools.org/remote-datasets/earth-relief.html}},
  note = {Cached gridline-registered dataset earth\_relief\_01d\_g.grd, derived from SRTM15+ V2.4 according to its stored metadata; the live documentation describes newer releases. Accessed October 3, 2026},
  year = {2026}
}

@article{kurganovPetrova2007,
  author = {Kurganov, Alexander and Petrova, Guergana},
  title = {A Second-Order Well-Balanced Positivity Preserving Central-Upwind Scheme for the {Saint-Venant} System},
  journal = {Communications in Mathematical Sciences},
  volume = {5},
  number = {1},
  pages = {133--160},
  year = {2007},
  doi = {10.4310/CMS.2007.v5.n1.a6}
}

@article{audusse2004hydrostatic,
  author = {Audusse, Emmanuel and Bouchut, Fran\c{c}ois and Bristeau, Marie-Odile and Klein, Rupert and Perthame, Beno\^{i}t},
  title = {A Fast and Stable Well-Balanced Scheme with Hydrostatic Reconstruction for Shallow Water Flows},
  journal = {SIAM Journal on Scientific Computing},
  volume = {25},
  number = {6},
  pages = {2050--2065},
  year = {2004},
  doi = {10.1137/S1064827503431090},
  url = {https://doi.org/10.1137/S1064827503431090}
}

@article{kirby2013dispersive,
  author = {Kirby, James T. and Shi, Fengyan and Tehranirad, Babak and Harris, Jeffrey C. and Grilli, St\'ephan T.},
  title = {Dispersive tsunami waves in the ocean: Model equations and sensitivity to dispersion and {Coriolis} effects},
  journal = {Ocean Modelling},
  volume = {62},
  pages = {39--55},
  year = {2013},
  doi = {10.1016/j.ocemod.2012.11.009},
  url = {https://doi.org/10.1016/j.ocemod.2012.11.009}
}

@misc{geoclawManning,
  author = {{Clawpack Development Team}},
  title = {Manning friction term},
  howpublished = {GeoClaw documentation, \url{https://www.clawpack.org/list/manning.html}},
  url = {https://www.clawpack.org/list/manning.html},
  note = {Accessed October 9, 2026}
}

@article{tozer2019srtm,
  author = {Tozer, B. and Sandwell, D. T. and Smith, W. H. F. and Olson, C. and Beale, J. R. and Wessel, P.},
  title = {Global Bathymetry and Topography at 15 Arc Sec: {SRTM15+}},
  journal = {Earth and Space Science},
  volume = {6},
  number = {10},
  pages = {1847--1864},
  year = {2019},
  doi = {10.1029/2019EA000658},
  url = {https://doi.org/10.1029/2019EA000658},
  note = {\href{https://doi.org/10.1029/2019EA000658}{doi:10.1029/2019EA000658}}
}

@book{abraham1988manifolds,
  author = {Abraham, Ralph and Marsden, Jerrold E. and Ratiu, Tudor},
  title = {Manifolds, Tensor Analysis, and Applications},
  edition = {2},
  series = {Applied Mathematical Sciences},
  volume = {75},
  publisher = {Springer},
  address = {New York},
  year = {1988},
  doi = {10.1007/978-1-4612-1029-0},
  url = {https://doi.org/10.1007/978-1-4612-1029-0}
}

@book{leveque2002finite,
  author = {LeVeque, Randall J.},
  title = {Finite Volume Methods for Hyperbolic Problems},
  series = {Cambridge Texts in Applied Mathematics},
  publisher = {Cambridge University Press},
  year = {2002},
  url = {https://www.cambridge.org/core/books/finite-volume-methods-for-hyperbolic-problems/97D5D1ACB1926DA1D4D52EAD6909E2B9}
}

@article{beuthe2015tides,
  author = {Beuthe, Mikael},
  title = {Tides on {Europa}: The membrane paradigm},
  journal = {Icarus},
  volume = {248},
  pages = {109--134},
  year = {2015},
  doi = {10.1016/j.icarus.2014.10.027},
  url = {https://doi.org/10.1016/j.icarus.2014.10.027}
}

@article{bermudez1994upwind,
  author = {Berm{\'u}dez, Alfredo and V{\'a}zquez, Mar{\'i}a Elena},
  title = {Upwind methods for hyperbolic conservation laws with source terms},
  journal = {Computers \& Fluids},
  volume = {23},
  number = {8},
  pages = {1049--1071},
  year = {1994},
  doi = {10.1016/0045-7930(94)90004-3},
  url = {https://doi.org/10.1016/0045-7930(94)90004-3}
}

@article{garcia2000source,
  author = {Garc{\'i}a-Navarro, Pilar and V{\'a}zquez-Cend{\'o}n, Mar{\'i}a Elena},
  title = {On numerical treatment of the source terms in the shallow water equations},
  journal = {Computers \& Fluids},
  volume = {29},
  number = {8},
  pages = {951--979},
  year = {2000},
  doi = {10.1016/S0045-7930(99)00038-9},
  url = {https://doi.org/10.1016/S0045-7930(99)00038-9}
}

\end{document}